\documentclass[usenatbib,useAMS,usegraphicx]{mn2e}
\input{journal.cls}

\newcommand{\lskip}{\vskip \baselineskip}
\newcommand{\nskip}{\lskip \noindent}
\newcommand{\halfskip}{\vskip 0.5\baselineskip}
\newcommand{\be}{\begin{equation}}
\newcommand{\ee}{\end{equation}}
\newcommand{\SWa}{\mbox{\rm SW1}}
\newcommand{\dAM}{\mbox{\rm dIGM}}
\newcommand{\uAM}{\mbox{\rm uIGM}}
\newcommand{\equref}[1]{(\ref{#1})}

\usepackage{hyperref}
\usepackage[table]{xcolor}
\usepackage{graphicx,epsfig}
\usepackage{verbatim}
\usepackage{syntonly}
\usepackage{txfonts}

\title[Simulating relativistic AGN jets II. Episodic activity]{Relativistic AGN jets II.
Jet properties and mixing effects for episodic jet activity}

\author[S. Walg et al.]{
S. ~Walg, $^{1,2}$\thanks{email: \url{s.walg@astro.ru.nl}}
A. ~Achterberg, $^1$
S. ~Markoff, $^2$
R. ~Keppens, $^3$
O. ~Porth $^4$
\\
$^1$Astronomical Institute, Radboud University Nijmegen,
                                   Heyendaalseweg 135, 6525 AJ Nijmegen, The Netherlands
\\
$^2$Astronomical Institute "Anton Pannekoek," University of Amsterdam,
                                   Science Park 904, 1098 XH Amsterdam, The Netherlands
\\
$^3$Centre for mathematical Plasma Astrophysics, Department of Mathematics, KU Leuven,
                                   Celestijnenlaan 200B, 3001 Heverlee, Belgium
\\
$^4$Department of Applied Mathematics, The University of Leeds, Leeds, LS2 9GT
}

\date{\today, accepted 6 February 2014}

\pagerange{\pageref{firstpage}--\pageref{lastpage}}
\volume{0000}
\pubyear{2014}

\begin{document}

\maketitle

\label{firstpage}

\begin{abstract}
Various radio galaxies show signs of having gone through episodic jet outbursts in the past. An example is
the class of double-double radio galaxies (DDRGs). However, to follow the evolution of an individual source in
real-time is impossible due to the large time scales involved. Numerical studies provide a powerful tool to
investigate the temporal behavior of episodic jet outbursts in a (magneto-)hydrodynamical setting.
We simulate the injection of two jets from active galactic nuclei (AGN), separated by a short interruption time.
Three different jet mo\-dels are compared. We find that an AGN jet outburst cycle can be divided into four
phases. The most prominent phase occurs when the restarted jet is propagating completely inside the hot and
inflated cocoon left behind by the initial jet. In that case, the jet-head advance speed of the restarted jet
is significantly higher than the initial jet-head. While the head of the initial jet interacts strongly with
the ambient medium, the restarted jet propagates almost unimpeded. As a result, the restarted jet maintains a
strong radial integrity. Just a very small fraction of the amount of shocked jet material flows back through
the cocoon compared to that of the initial jet and much weaker shocks are found at the head of the restarted
jet. We find that the features of the restarted jet in this phase closely resemble the observed properties of a
typical DDRG.
\end{abstract}

% \nokeywords


\section{Introduction}
\label{sec:Introduction}

\begin{figure}
%\includegraphics[clip=false,width=0.5\textwidth,angle=0]{pics/b1545-321.eps}
% !!! DOWNLOADED FROM http://outreach.atnf.csiro.au/images/astronomical/b1545-321.html
\includegraphics[clip=false,width=0.5\textwidth,angle=0]{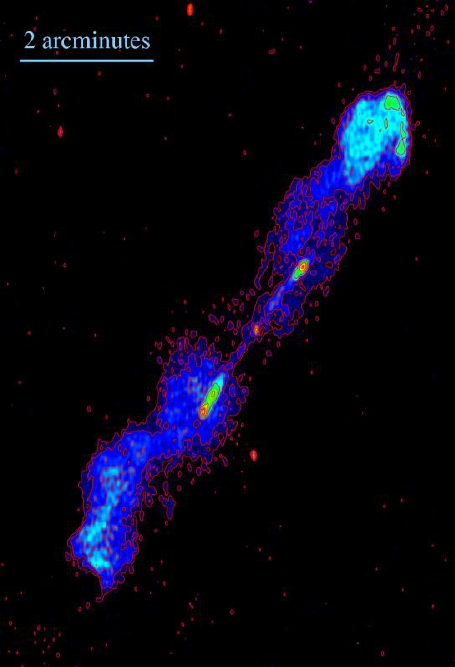}
\caption{Non-thermal (synchrotron) radio emission of the Double-Double Radio Galaxy \mbox{PKS B1545-321},
observed at 1384 MHz and 2496 MHz with the ATCA. Two prominent outer radio lobes and two less prominent inner
lobes are visible. The total size of the source is \mbox{$\sim$ 1 Mpc}. The North-East part of the source is
pointed away from the observer. Credits: \citet{Saripalli2003}. 
}

  \label{fig:DDRG}
\end{figure}

Over the last decade, a new class of radio sources called {\em Double-Double Radio Galaxies} (DDRGs) has been
identified (e.g. \citealt{Schoenmakers2000}; \citealt{Saikia2009}). These sources can be characterized by a pair
of double hotspots and/or radio lobes driven by the same SMBH. Some rare examples have even been reported where
three distinct pairs of radio lobes are seen on both sides of the AGN. In that case the source is referred to as
a Triple-Double Radio Galaxy \citep[e.g.][]{hota2011}. Figure \ref{fig:DDRG} shows the radio map for the source
\mbox{{\em PKS B1545-321}} (B1545-321 hereafter), a typical example of a DDRG
(\citet*{Saripalli2003}; \citet{Safouris2008}). Of the (roughly 20) DDRGs that are known to date,
the distance of the inner radio lobes to the central engine range from as close as 14 pc as in J1247+6723
\citep*{Saikia2007}, up to several hundreds of kpc as in J1835+6024 \citep{Lara1999}. The distance of the outer
lobes to the central engine is usually much larger, of the order of a few Mpc.

\halfskip
Jet properties such as stability and the integrity of the transverse (radial) structure, the impact of the jet on
the intergalactic medium (IGM) and the closely related jet-head advance speed depend strongly on the properties
of the ambient medium through which the jets propagate. If an AGN undergoes multiple cycles of activity, it is
capable of drastically changing this ambient medium. Multiple pairs of radio lobes and hotspots within the same
radio galaxy are strong indications for the occurrence of episodic jet eruptions in AGNs
(see for example
\citealt[][]{Saripalli2003};
\citealt[][]{Safouris2008};
\citealt[][]{Konar2013a};
\citealt[][]{Konar2013b};
\citealt[][]{Konar2013c};
\citealt[][]{Konar2013d}). In some cases the interruption time, that is the time between two subsequent jet
eruptions, for DDRGs is short compared to the duration of the initial jet eruption, as recent studies suggest
(\citealt[][]{Konar2013b}; \citealt[][]{Konar2013c}). Then, the
signatures from both eruptions (mainly in the form of synchrotron radiation) are observed simultaneously.
However, if the interruption time is comparable with, or larger than the duration of the initial jet eruption,
the inflated hot cocoon (a mixture of shocked IGM and shocked jet material, hereafter `$\dAM$') from the first
eruption has the chance to expand and the non-thermal electrons have the chance to cool significantly before the
next eruption begins. In that case, the radiation from the relativistic leptons produced in the initial jet
eruption might no longer be observable, whereas the conditions of the disturbed medium still significantly differ
from those of the undisturbed ambient medium (hereafter `$\uAM$').
\begin{comment}
An example of where this effect potentially plays a crucial role is in the division between the FRI/FRII
morphology: the difference in morphology might be caused by the change of the direct environment of the source on
kpc-Mpc scales as a result of earlier jet eruptions (also see \citealt[][]{Chon2012}).
\end{comment}
Having a better understanding of episodic AGN jet activity, the time scales involved and the effect of episodic
outbursts on the ambient medium will contribute to a more global understanding of AGN jets and galaxy evolution
in general.

%  \subsection{Scientific motivation}
\lskip
There are a number of reasons for studying episodic AGN jet behavior. For an AGN that launches a jet into an
$\uAM$, jet properties (such as jet-head advance speed, stability and transverse structural integrity, or the
dynamics of the hotspots, cocoon and back-flow of shocked jet material), are expected to be quite different from
jets in a source that shows multiple outbursts. This is even true when the intrinsic characteristics such as
power, the direction into which the jet is injected, opening angle, injection spectral index, etc. are equal. It
is impossible to directly follow AGN jet eruption cycles due to the enormous length/time scales involved (up to
hundreds of Myr, see for example \citealt*{McNamara2007}; \citealt{Wise2007} or \citealt*{McNamara2012}).
Therefore, observational evidence for the existence of episodic AGN jet cycles in an individual source can only
indirectly be inferred from synchrotron spectral ages and morphology, such as the distances of different
synchrotron-emitting regions to the central engine. Fortunately, hydrodynamical (HD) simulations of relativistic
AGN jets provide a tool for calculating the (strong non-linear) large-scale behavior of these jet flows and
their surrounding ambient medium, offering a powerful method to ultimately help model the observations. The aim
is then to identify the characteristic features that are seen in the different phases of a jet outburst in the
simulation and compare them with the observed signatures in the emitted non-thermal radiation from such a
radio source.

Numerical studies that are related to time-dependent/episodic jet behavior go back to the study of
\citet[][]{Wilson1984} who simulates jets with a sinusoidally varying jet speed.
\begin{comment}
Although a number of aspects regarding temporal jet behavior are treated, this study is not aimed at
investigating the properties of a typical DDRG.
\end{comment}
\mbox{\citet[][]{Clarke1991}} simulate restarting 2-dimensional magnetohydrodynamic \mbox{(2D MHD)}
non-relativistic jets. In powerful radio galaxies where jets show typical Lorentz factors of a few up to
$\sim 50$, relativistic effects need to be taken into account in order to simulate a realistic scenario.
\citet{Chon2012} perform a \mbox{3D HD} simulation of a restarting jet in the FRII radio galaxy Cygnus A. 
Their paper focusses on an X-ray cavity observed near the parent galaxy of the radio source Cygnus A that is
believed to be created by a jet event that took place ∼ 30 Myr ago. By simulating the X-ray emission at a late
time during the second jet eruption, they find that the regions containing jet material from the first eruption
show up as X-ray cavities, similar to those observed in Cygnus A. Although a relevant case study for Cygnus A,
the intermediate phases leading to its current state, as well as the jet dynamics itself are not treated.
Finally \citet*{Mendygral2012} study restarting 3D non-relativistic MHD jets injected into the external medium
of a galaxy cluster to investigate the influence of this intra-cluster medium (ICM) on the morphological
evolution of the jets and their radio lobes. These simulations mainly probe the large-scale characteristics of
the resulting radio lobes and X-ray cavities.

All these studies consider non-relativistic jets that are homogeneous in the radial direction. An exception is the
work of \citealt{Mendygral2012}, who include a radially varying electromagnetic field. 
However, AGN jets show strong signs of stratification transverse to their direction of propagation (i.e. radial
stratification). The observations favor a jet consisting of a low-density and fast-moving spine, surrounded by a
denser and slower moving region called the jet sheath (see for example
\citealt*{Sol1989};
\citealt{Giroletti2004};
\citealt*{Ghisellini2005};
\citealt{Gomez2008}).
In \mbox{\citet{Walg2013}}, hereafter $\SWa$, we presented the study of the evolution of relativistic axisymmetric
jets with three different transverse jet profiles carrying angular momentum, for a steady jet driven by a
continuous inflow.

\section{Main focus of this research}
\label{sec:MainFocus}
We present the first special relativistic simulations of episodic outbursts of AGN jets. These simulations allow
us to accurately study the jet dynamics, as well as the amount of shocked back-flowing jet material at the
jet-head.
To that end, we simulate two distinct episodes of jet activity. We keep track of the constituents of both jets.
The aim here is to understand the processes that lead to the morphology of a DDRG such as B1545-321. In the
follow-up paper, these same simulations will be used to study synchrotron emission at the various stages of an
episodic jet event. Our choice of parameters is representative for a typical powerful radio galaxy and is
representative for a typical DDRG. We will focus on the following points:

\begin{itemize}
\item How does jet stratification change during an activity cycle;
\item How does the propagation of the restarted jet in the cycle differ from the first jet, and what determines
the jet-head advance speed;
\item How rapidly do the strong shocks at the jet-head fade after the first jet has been turned off;
\item How does mixing between the different constituents (ambient medium, spine material, sheath material) take
place and how is the amount of mixing influenced by episodic activity. In particular: to what amount does
material from the first and the second jet mix.
\end{itemize}
\nskip
All these features that follow from the simulations will add to our understanding and interpretation of the rich
diversity and large-scale structures that are observed in giant radio galaxies, such as DDRGs.

\lskip
The outline of this paper is as follows: in Section \ref{sec:TBG} we present the background theory. In
Section \ref{sec:Method} we discuss the method, numerical schemes and the parameter regime. In Section
\ref{sec:Results} we describe the different simulations and their results. Discussion and Conclusions can be
found in Sections \ref{sec:Discussion} and \ref{sec:Conclusions}.

\section{Theoretical background}
\label{sec:TBG}

  \subsection{Relation between the DDRG synchrotron emission and the dynamics of the jet flow}
\label{subsec:Rel_DDRG-jet}

The intensity of the synchrotron emission that is emitted from an extragalactic radio source (at a certain
acceleration site) depends on [1] the rate of injection of relativistic particles, [2]
the frequency $\nu$ at which the source is observed and [3] the synchrotron cooling time
$\tau_{\rm synchro}(\nu)$ at that frequency. In addition, adiabatic cooling due to source expansion affects the
synchrotron intensity at all frequencies. The time-scale for adiabatic cooling, $\tau_{hydro}$, is closely
related to the hydrodynamical properties (and time-scales) of the gas.
\begin{comment}
The resulting cooling time $\tau_{\rm cool}$ for a system where both cooling mechanisms are at work can be
written as:

\be
\tau_{\rm cool} \simeq \frac{\tau_{\rm synchro} \tau_{\rm hydro}}{\tau_{\rm synchro} + \tau_{\rm hydro}} \; .
\ee
\end{comment}
There are two limiting cases, namely:
\begin{description}
\item \mbox{$\tau_{\rm hydro} \gg \tau_{\rm synchro}$}, in which case the energy losses are dominated by the
synchrotron cooling;
\item \mbox{$\tau_{\rm hydro} \ll \tau_{\rm synchro}$}, in which case the energy losses are dominated by the
adiabatic cooling (expansion losses).
\end{description}

In this paper, the dynamical time scales, such as duration of the active outbursts phases or the interruption
times, are \mbox{$\sim 10^6 - 10^7$ yr}. These time-scales are short compared to the synchrotron cooling time at
radio frequencies of \mbox{$\sim 1 - 2$ GHz}, which are typical frequencies at which DDRGs are observed (see for
example \citealt{Harwood2013}). Therefore, the energy losses of the electrons responsible for the synchrotron
emission for the radio galaxies that we are considering are dominated by the effect of the expansion of the gas.

A typical DDRG is believed to be the result of a restarting jet propagating inside a disturbed medium $\dAM$ of an
earlier eruption that took place a relatively short time ago (interruption time much less than lifetime of an
individual jet). Moreover, since the radio lobes and/or hotspots of the initial jet eruption are still visible,
either the outer radio lobes and hotspots are still being fed by fresh material from the initial jet (as is
assumed to be the case for the DDRG J1835+6204), or the initial jet has disappeared very recently as is believed
to be the case for B1545-321. We will consider a similar situation for the jet models in this paper. We will
follow the restarting jet up to and including the propagation outside of the initial cocoon, in the $\uAM$.

%\begin{comment}
{\renewcommand{\arraystretch}{1.3}
\begin{table}
\caption{List of computer normalization units (or characteristic quantities) that we chose for these
simulations in cgs units. These characteristic quantities are the same as in $\SWa$ and apply throughout this
paper.}
%\rowcolors{1}{white}{lightgray}
%\begin{minipage}{0.5\textwidth}
%\centering
%  \begin{center}
    \begin{tabular}{ l  c  c } \hline
       {\bf Char. quantities} & {\bf symbol} & {\bf cgs units}                              \\
       \hline
                      Number density   & $n_{\rm ch}$ & $10^{-3}\; {\rm cm}^{-3}$                    \\
       
                            Pressure   & $P_{\rm ch}$ & $1.50 \times 10^{-6}\; {\rm erg \; cm}^{-3}$ \\
      
                         Temperature   & $T_{\rm ch}$ & $1.09 \times 10^{13}\; {\rm K}$              \\
       \hline
    \end{tabular}
    \label{tab:charvars}
%  \end{center}
%\end{minipage}
\end{table}
}
%\end{comment}

  \subsection{Outburst cycle time scales for DDRGs}
\label{subsec:OutburstCycleTime}

The age of a radio galaxy is often inferred from the synchrotron spectral age of its radio lobes or its plumes
\citep[e.g.][]{Harwood2013}.
\begin{comment}
Two particular subclasses of radio sources, named Gigahertz Peaked-Spectrum radio sources (GPS) and Compact
Steep-Spectrum radio sources (CSS) are relatively small and young sources. These sources might be the
predecessors of the larger FRI or FRII radio sources. Based on the spectral properties of these sources, the
synchrotron lifetime of the relativistic electron population is estimated to lie in the range
\mbox{$10^3 - 10^6$ yr}.
Though an exact relation between the synchrotron age and the age of the sources is unknown, they could be mutually
consistent (see for instance \citealt{ODea1998}). 
\end{comment}
In this way, some of the larger sources are estimated to be of the order of \mbox{$10^6-10^8$ yr} old (see for
example
\citealt{Jamrozy2008};
\citealt{Konar2008};
\citealt{ODea2009} or
\citealt*[][]{Saikia2009}).

Recent studies suggest that, for BH systems with an accretion rate well below the Eddington limit,
\mbox{$\dot{M}_{\rm BH} \ll \dot{M}_{\rm edd}$}, accretion and jet formation behave in a similar fashion,
scalable by BH mass. An indication for this scaling was found by \citealt{McHardy2006}, where they show
a mass dependence of characteristic time scales for the variability of the emission produced near black holes
of black hole binaries (BHBs) and AGNs. Another strong indication for mass scaling accretion physics is the
observed correlation between the luminosities in X-ray and Radio frequencies for these systems. A number of
authors
(e.g. \citealt{Corbel2000};
\citealt{Corbel2003};
\citealt*{Merloni2003};
\citealt*{Falcke2004};
\citealt*{Kording2006} and
\citealt{Plotkin2012})
have studied the relationship between X-ray luminosity ($L_{\rm{X}}$), radio luminosity ($L_{\rm{R}}$) and
BH mass ($M_{\rm{BH}}$) for BH systems with a low accretion rate in BHBs, as well as in SMBHs in the
centre of active galaxies. They have shown a relationship between $L_{\rm{X}}$, $L_{\rm{R}}$ and $M_{\rm{BH}}$
that holds over many orders of magnitude in BH mass, ranging from stellar-mass BHs with a typical mass of
\mbox{$M_{\rm BH} \sim 10 M_{\odot}$} up to the largest SMBHs with a typical mass of
\mbox{$M_{\rm BH} \sim 10^9 M_{\odot}$}. This relation defines a plane in three-dimensional
\mbox{$(\log L_{\rm{X}},\log L_{\rm{R}},\log M_{\rm{BH}})$} parameter space, called the \textit{fundamental
plane of BH accretion}.

%\begin{comment}
{\renewcommand{\arraystretch}{1.3}
\begin{table*}
\caption{Global parameters of the ambient medium and the jet at the jet inlet as used in the models $H2$, $I2$,
$A2$. From left to right they are: kinetic luminosity ($L_{\rm jt}$), number density ($n$), Lorentz factor
($\gamma$), azimuthal velocity ($V_{\phi}$) and polytropic index ($\Gamma$) used to setup the transverse (radial)
pressure profile of the jets. In case of model $H2$, the jet is homogeneous in the transverse direction and is
described by single-valued quantities. The parameters for models $I2$ and $A2$ are initialized separately for
spine (denoted as "sp" in the table) and sheath (denoted as "sh" in the table). In case of model $I2$ and $A2$,
the pressure varies smoothly in the radial direction.}

\begin{tabular}{l c c c c c} \hline
{\bf Models} &
$L_{\rm jt} \; [10^{46}  \; \rmn{erg \; s^{-1}} ]$ &
$n         \; [10^{-6}  \; \rmn{cm^{-3}}       ]$ &
$\gamma$                                             &
$V_{\phi}  \; [10^{-3}  \; c]$                       &
$\Gamma$                                             
\\

\phantom{}                                           &
\begin{tabular}{c c c} \hline
sp & $|$ & sh
\end{tabular}                                        &
\begin{tabular}{c c c} \hline
sp & $|$ & sh
\end{tabular}                                        &
\begin{tabular}{c c c} \hline
sp & $|$ & sh
\end{tabular}                                        &
\begin{tabular}{c c c} \hline
sp & $|$ & sh
\end{tabular}                                        &
\begin{tabular}{c c c} \hline
sp & $|$ & sh
\end{tabular}                                        
\\ \hline

%\rowcolor{lightgray}
      $\mathbf{H}$ {\bf (homogeneous)} & 3.82 & 4.55    & 3.11  & 0.0 & 1
      \\
      $\mathbf{I}$ {\bf (isothermal)} & 
                                         \begin{tabular}{c c c}
                                             1.82 & \phantom{} & 3.35
                                         \end{tabular}
                                         &
                                                $P/\rho = $ constant
                                            & 
                                            \begin{tabular}{c c c}
                                                6.0 & \phantom{} & 3.0
                                            \end{tabular}
                                              &
                                              \begin{tabular}{c c c}
                                                  1.0 & \phantom{} & 1.0
                                              \end{tabular}
                                                  &
                                                  \begin{tabular}{c c c}
                                                      5/3 & \phantom{} & 5/3
                                                  \end{tabular}
                                                      \\
%  \rowcolor{lightgray}
      $\mathbf{A}$ {\bf (constant density)} & 
                                         \begin{tabular}{c c c}
                                             0.44 & \phantom{} & 3.39
                                         \end{tabular}
                                         &
                                      \begin{tabular}{c c c}
                                          1.0 & \phantom{} & 5.0 
                                      \end{tabular}
                                          & 
                                          \begin{tabular}{c c c}
                                              6.0 & \phantom{} & 3.0
                                          \end{tabular}
                                              &
                                              \begin{tabular}{c c c}
                                                  1.0 & \phantom{} & 1.0
                                              \end{tabular}
                                                  &
                                                  \begin{tabular}{c c c}
                                                      5/3 & \phantom{} & 5/3
                                                  \end{tabular}
                                                      \\
      {\bf External medium} & - & $1.0\times 10^{3}$ & - & - & 5/3 \\
      \hline
    \end{tabular}

    \label{tab:parameters}

\end{table*}
}

Many BHBs are observed to go through outburst cycles (e.g. \citealt{Rodriguez1999}; \citealt{Fender2002}). It is
generally believed that AGNs also go through outburst cycles, but it is unclear whether these cycles are driven
by the same physical mechanisms. A scenario that can explain the origin and the specific morphology of DDRGs
is given by \citet*{Liu2003} who suggest an inspiraling SMBH binary as a result of two merging galaxies to be the
cause. In their model, the secondary SMBH slowly sinks towards the centre of mass, causing a gap in the
inner accretion disc to occur. In this way, the jet formation of the primary SMBH is temporarily stopped. When
the gap is refilled by material from the outer accretion disc (on a viscous time-scale of \mbox{$\sim 1$ Myr}),
jet formation is restarted. Cycle times for accreting SMBHs are typically long (\mbox{$\sim 10^6 - 10^8$ yr}).
These large time-scales suggest that, at least in sources like DDRGs, the observed morphology of the source is
essentially a snapshot of an ongoing eruption event.

\section{Method}
\label{sec:Method} 

  \subsection{The jet models, the parameters and initial conditions}

Consider a central engine in the nucleus of a radio galaxy that undergoes two or more subsequent jet eruptions.
If the different episodes are triggered by the same mechanism (for example, a large in-falling gas cloud), it is
reasonable to assume that the typical initial jet properties near the central engine, as for example mass
density, bulk outflow velocity, rotation, etc., should be approximately similar for the subsequent jet eruptions
(also see \citealt*[][]{Konar2013a}; \citealt[][]{Konar2013d}). In this paper we will assume all the jet
parameters to be equal for the two subsequent jet eruptions.

We employ {\small MPI-AMRVAC} \citep{Keppens2012} and simulate the jets with a special relativistic
hy\-dro\-dyna\-mical (SRHD) module. We use the spatial Harten-Lax-van Leer Contact (HLLC) solver
(\citealt*{Toro1994}; \citealt*{Migone2005}) for the three jet models, in combination with a three-step
Runge-Kutta time-discretization scheme and a Koren limiter \citep{Koren1993}. 

The jets are cylindrically symmetric ($2.5D$) with the $Z-$axis defined along their axis. The jet flows are
created by injecting material into the computational domain through the boundary cells at the $Z = 0$ axis,
between \mbox{$R = 0$} and \mbox{$R = R_{\rm jt} =$ 1 kpc}, which we refer to as the {\em jet inlet}. Except for
the cells involved in injecting the jet material (during the active phase of jet injection), all other cells in
the lower boundary are free outflow boundaries. In case of the structured jets, we choose the radius of the
spine equal to \mbox{$R_{\rm s} = R_{\rm jt}/3$}.

The size of the computational domain is $(120 \times 480)$ kpc$^2$, comparable to the size of either one of the
jets in \mbox{B1545-321} (see \mbox{Figure \ref{fig:DDRG}}). The basic resolution is \mbox{$(120 \times 480)$}
grid cells, and we allow for 3 additional refinement levels, resulting in an effective resolution of
\mbox{$(960 \times 3840)$} grid cells. The jet is resolved by 8 grid cells across the jet radius. Therefore, we
can resolve details down to \mbox{$(125 \times 125)$ pc$^2$}.

For the polytropic index $\Gamma_{\rm eff}$ of the gas we work with the Mathews approximation for the Synge EOS
of the gas \citep{Blumenthal1976}. In this approximation, the gas pressure $P$ is defined by the closure
relation:

\be
        P = \frac{1}{3} \left( e - \frac{\rho^2}{e} \right) \; ,
            \label{eq:closurerelation}
\ee
with $e = e_{\rm th} + \rho$, the internal energy density consisting of the rest-mass density $\rho$ and thermal
energy density $e_{\rm th}$. We employ units where $c = 1$. This approximation gives an accurate interpolation
between a classically 'cold' gas with an adiabatic index \mbox{$\Gamma_{\rm eff} = 5/3$} and a relativistically
'hot' gas with \mbox{$\Gamma_{\rm eff} = 4/3$}. We define the transition from cold to relativistically hot to
occur when the internal energy per particle is equal to the rest-mass energy per particle, i.e.
\mbox{$k_{\rm B} T = m_{\rm p} c^2$}, resulting in \mbox{$\Gamma_{\rm eff} = 1.417$}.

The jet models that are used in this paper are the same as those used in $\SWa$. They include a transverse
homogeneous jet ($H$) and two jets with a transverse spine--sheath jet structure that carry angular momentum. At
the jet inlet, radial force-balance is maintained along the jet cross-section. The condition of radial
force-balance, together with the azimuthal velocity profile, $v_{\phi}(R)$ determine the pressure profile,
$P(R)$, across the jet cross-section. However, the azimuthal velocity at the length scales that we are considering
are thought to be small compared to the poloidal velocity (\mbox{$v_{\phi} \ll v_{\rm Z}$}), so it has a
negligible influence on the dynamics of the jets.

One jet with radial structure is set up using an isothermal equation of state ($I$), which we
denote as the {\em isothermal jet}. The other uses a constant, but different density for spine and sheath ($A$),
which we denote as the (piecewise) {\em isochoric jet}. To distinguish between the steady case scenario
(denoted with an index `1'), and the episodic scenario, we refer to the models in this paper as $H2$, $I2$ and
$A2$ respectively. The isochoric jet $A2$ and the isothermal jet $I2$ consist of a jet spine--sheath structure
that is characterized by having a different bulk Lorentz factor for jet spine and jet sheath. The isochoric jet
$A2$ is initiated and injected with a constant, but different mass density for the jet spine and the jet sheath.
The isothermal jet $I2$ is initiated and injected with a constant temperature across the entire cross-section.

The jet models are based on realistic values for mass density \mbox{$\rho = m_{\rm p}n$}, with $m_{\rm p}$ the
proton mass, $n$ the number density and temperature $T_{\rm am}$ of a typical ICM environment (see e.g.
\citealt{Dave2001}; \citealt{Dave2010} and \citealt{Kunz2011}). Moreover, jet bulk
Lorentz factors $\gamma_{\rm jt}$ are used that are typical for jets driven by SMBHs at these length scales. The
density ratio \mbox{$\eta_{\rm R} = n_{\rm jt}/n_{\rm am}$} between the jet and the ICM are deduced from the jet
luminosity $L_{\rm jt}$.
\begin{comment}
Finally, the jet azimuthal velocity $v_{\phi}$ (that is carried by the isothermal and the isochoric jet) is
estimated, based on the relativistic MHD properties near the central engine and further out.
\end{comment}
In the initial setup, all jets are in pressure equilibrium with their surrounding at the interface between the
jet and the ambient medium. \mbox{Tables \ref{tab:charvars} and \ref{tab:parameters}} show the parameters used
in these models. For a detailed discussion on the set up of the radial jet profiles, we refer the reader to $\SWa$.

  \subsection{Quantifying mixing for multiple constituents}
\label{subsec:Mixing}

\begin{comment}
To understand how material is advected and/or mixed with other constituents, this material must be traced.
Fluid-tracing becomes especially important when dealing with multiple constituents, as is the case for jets with
a transverse spine--sheath structure and/or for episodic jets. To that end
\end{comment}
We employ {\em tracers}, \mbox{$\theta_{\rm A}(t,\bmath{r})$}, that are passively advected by the flow from cell
to cell in the numerical grid employed in the simulations. We initiate them as follows:
\begin{description}
\item {\em H2}: for the homogeneous jet with two eruptions we use two tracers $\theta_{i}$, with $i = 1$ for
the material involved in the initial jet eruption and $i = 2$ for the subsequent one. Jet material is initialized
as $\theta_{i} = +1$ and ambient medium material as $\theta_{i}=0$.
\item {\em A2 and I2}: these models simulate the case of episodic jets with a transverse spine--sheath structure.
For these jets we use four tracers, $\{ \theta_{i}^{\rm sp}, \theta_{i}^{\rm sh} \}$, with again $i = 1$ for
material involved in the initial eruption and $i = 2$ for the subsequent one. The tracers $\theta_{i}^{\rm sp}$
are initialized as +1 for spine material and as 0 elsewhere. Equivalently, the tracers $\theta_{i}^{\rm sh}$ are
initialized as +1 for sheath material and 0 elsewhere
\footnote{The initialization differs slightly from the one used in $\SWa$, where the minimum tracer value was
chosen \mbox{$\theta_{\min} = -1$}, instead of 0 in this paper.}.
\end{description}
\begin{comment}
As the tracers are advected, mixing, as well as numerical discretization effects will affect tracer values within
a grid cell. Generally, one obtains values between \mbox{$0 \le \theta_{\rm A}(t,\bmath{r}) \le +1$}.
\end{comment}
We will interpret the tracer value $\theta_{\rm A}(t,\bmath{r})$ in a certain grid cell to be equal to the mass
fraction $\delta_{\rm A}(t,\bmath{r})$ of constituent $A$ in that grid cell (and similar for constituent $B$).
\begin{comment}
In order to interpret observations of the non-thermal radio emission coming from sources such as DDRGs, it is
important to understand the contribution from each jet component to the total emission. In HD simulations,
keeping track of different jet components can be achieved by advecting tracers with the gas flow. The tracer
value $\theta_{\rm A}$ of constituent $A$ in a certain grid cell, and the mass fraction $\delta_{\rm A}$ of that
constituent in that grid cell have a one-to-one correspondence.

When dealing with more than two constituents, the mass fraction of a number of constituents in a grid cell does
not always give a clear view on the extent to which these constituents have mixed, especially when dealing with
more than two constituents. Mixing can be particularly useful for studying the entrainment of one fluid or gas by
another, or studying the structural integrity of a flow, such as a transverse structured jet.
\end{comment}

In $\SWa$, we performed a quantitative analysis of mixing. We introduced two practical mixing quantities, the first
called {\em absolute mixing}, denoted as $\Delta$ and the second called {\em mass-weighted mixing}, denoted as
$\Lambda$. These two quantities can compactly be written as:

\begin{equation}
\mathcal{M}_{\rm J} = 1 - \left| \frac{\delta_{\rm A} - \mu_{\rm J} \delta_{\rm B}}{\delta_{\rm A} +
\mu_{\rm J} \delta_{\rm B}} \right| \; ,
\end{equation}
with \mbox{$\mathcal{M}_{\rm J} = \{\Delta,\Lambda\}$} the absolute mixing and mass-weighted mixing respectively,
where we dropped the notation for the time- and space dependence, $(t,\bmath{r})$. Moreover,
\mbox{$\mu_{\rm J} = \{1,M_{\rm A}/M_{\rm B}\}$}, with $M_{\rm A}$ and $M_{\rm B}$ the total masses of
constituents $A$ and $B$ contained in the total computational volume respectively. With this definition for
the amount of mixing, \mbox{$\mathcal{M}_{\rm J} = 0$} means the constituents $A$ and $B$ have not mixed at all,
whereas \mbox{$\mathcal{M}_{\rm J} = 1$} means the constituents have fully mixed. We refer the reader to $\SWa$
for a more detailed discussion on tracer advection and mixing.

\begin{comment}
The absolute mixing $\Delta$ uses the mass fraction of constituents $A$ and $B$ in a volume element directly.
It means that for \mbox{$\Delta = 0$}, the constituents have locally not mixed at all, while for
\mbox{$\Delta = 1$}, the mass fractions of the constituents $A$ and $B$ in that grid cell are equal.

The mass-weighted mixing $\Lambda$ divides the mass fraction of constituent $A$ in a grid cell by the total
mass $M_{\rm A}$ contained in the total volume, and similar for constituent $B$. If the constituents $A$ and $B$
would settle into a completely homogeneous mixture in some volume $V$, then the mass-weighted mixing yields a
value of $\Lambda = 1$, everywhere in $V$. On the other hand: when \mbox{$\Lambda = 0$} in a grid cell this
means that {\em in this cell} constituents have not mixed at all.
\end{comment}

  \subsection{Time scales of the various phases of the jet eruptions}
\label{subsec:TimeScales}

%%%%% Jet inlet fraction %%%%%
\begin{figure}
%\begin{center}
%$
%\begin{array}{cc}
%\includegraphics[clip=true,trim=2cm 0cm 0cm 2.5cm,width=0.5\textwidth,angle=0]{pics/InletFraction/InletFraction.eps} \\
\includegraphics[clip=true,trim=2cm 0cm 0cm 2.5cm,width=0.5\textwidth,angle=0]{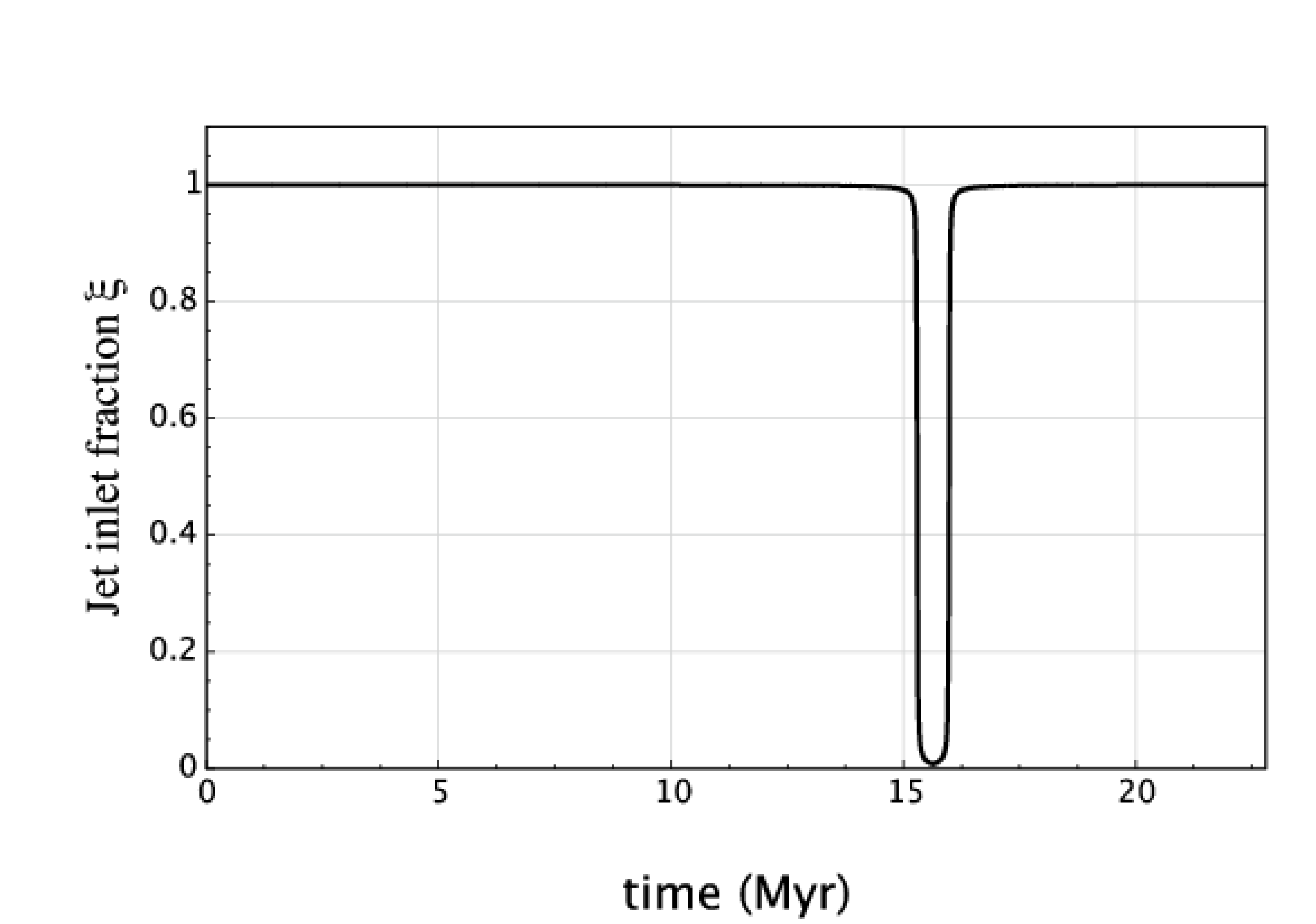} \\
%\end{array}
%$
%\end{center}
\caption{Jet inlet fraction showing the activity of the central engine that is driving the jets. The poloidal
velocity of the jets $v_{z}(t)$ is determined by \mbox{$v_{z} = \xi(t) \times V_{z}$}, where the parameter
$V_{z}$ is a fixed number and follows from the Lorentz factor of that jet component. At $t$ = 15.3 Myr, the
central engine switches off and at $t$ = 16.0 Myr, a subsequent jet is injected into the system for another
6.8 Myr. The simulation is stopped at $t$ = 22.8 Myr.}
  \label{fig:Inlet}
\end{figure}

%%%%% Figures of the jet-head propagation %%%%%
\begin{figure}
%\begin{center}
%$
%\begin{array}{cc}
%\includegraphics[clip=true,trim=0cm 0cm 0cm 2.5cm,width=0.5\textwidth,angle=0]{pics/jhas/jhasA2H2I2.eps} \\
\includegraphics[clip=true,trim=0cm 0cm 0cm 2.5cm,width=0.5\textwidth,angle=0]{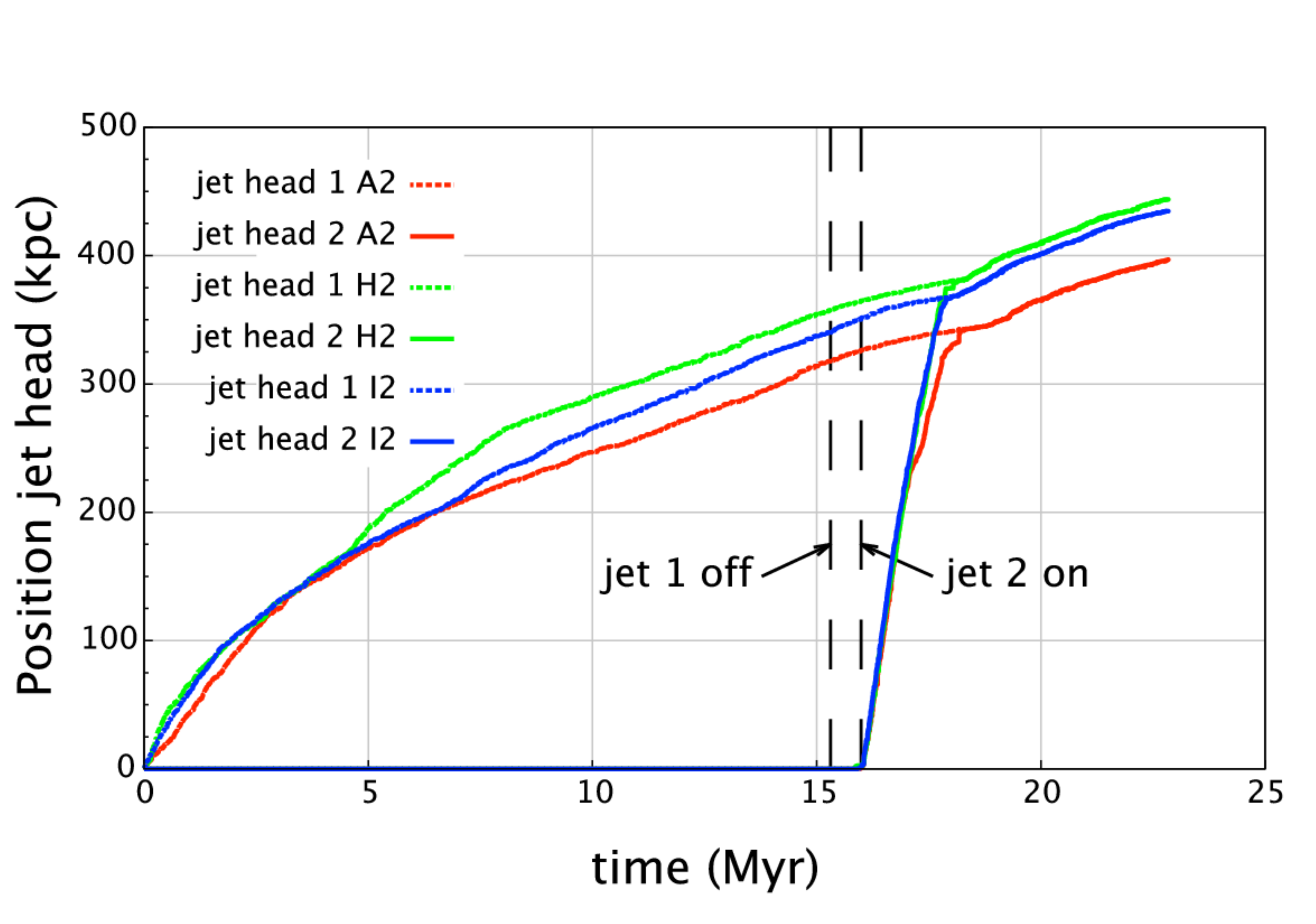} \\
%\end{array}
%$
%\end{center}
\caption{Jet-head propagation for two subsequent episodic jet outbursts for models $A2$, $H2$ and $I2$. As
with the jet models in $\SWa$, the initial jets show a start-up phase of \mbox{$\sim 3$ Myr}, where the jet-head
advance speed is slightly larger than after that phase. At \mbox{$t$ = 15.3 Myr}, the initial jets are switched
off. However, the left-over jet continues to propagate towards the termination shock. At \mbox{$t$ = 16.0 Myr}, a
restarting jet is injected into the system. It propagates significantly faster through the remnant cocoon, until
it hits the edge of the remnant cocoon. At that point, it continues to propagate in the same fashion as the
initial jets.}
  \label{fig:JHAS-2}
\end{figure}

The age of a typical DDRG is estimated to be of the order of \mbox{$10^7 - 10^8$} yr with a total length of
\mbox{$\sim 1$ Mpc}, as for example in the case of B1545-321 (see \citealt{Safouris2008}). Moreover, the
interruption time between the two jet events is usually short, only a few percent of the duration of the first
jet eruption. In the simulations in this paper, we choose a similar setup, but consider a length/time scale that
is roughly a factor of 2 smaller than that of B1545-321. Choosing the total simulation time equal to that in
$\SWa$ (\mbox{$t_{\rm tot} = 22.8$ Myr}) [1] allows for a fair comparison between the two studies, [2] captures
the general behavior of an episodic AGN jet event and [3] does not exhaust computational resources.
In order for the initial jet/cocoon to reach a typical length of a few hundred kpc, we take the initial eruption
time \mbox{$t_{\rm jet_1} = 15.3$ Myr} \mbox{( = $2/3 \times t_{\rm tot}$)}. Then we choose an interruption time
of \mbox{$t_{\rm int} = 0.045 \times t_{\rm jet_1} = 0.68$ Myr}. With this choice, we find that the length of the
restarted jet is approximately 1/3 of the cocoon at the moment that the first jet disappears, in
agreement with the morphology of B1545-321 (see \mbox{Figure \ref{fig:DDRG}}). Finally, we inject the second
jet for the remaining \mbox{$t_{\rm jet_2} = 6.8$ Myr}. We stop the simulations at the point where the second
jet-head has completely traversed the $\dAM$, so that the front end of the jet is propagating in the $\uAM$, but
before the jet runs out of the computational domain.

\begin{comment}
The exact details of how an AGN jet is shut down or turned back on are unknown. It might be expected that this
proceeds in a somewhat smooth fashion. Moreover, introducing sharp cut-offs for the jet injection results in
numerical errors.
\end{comment}
In order to model a smoothly varying injection of jet material at the jet inlet, we introduce
a time-dependent poloidal jet bulk velocity $v_z(t)$ by multiplying the axial velocity $V_z$ with the jet inlet
fraction $\xi(t)$, so that at a given time: \mbox{$v_z(t)=\xi(t) \times V_z$}. The jet inlet fraction $\xi(t)$
is plotted in \mbox{Figure \ref{fig:Inlet}}. For $\xi(t)$ we use the following form:

\be
\xi(t) = \left\{
\begin{array}{ll}
\tau_1(t) + 1 - \tau_1(t=t_0) & \mbox{for} \; t \le t_{\rm c} \\
\tau_2(t) + 1 - \tau_1(t=t_{\rm tot}) & \mbox{for} \; t > t_{\rm c} \; ,
\end{array}
\right.
\label{eq:InletFraction}
\ee
with $t$ in Myr. Here $t_{\rm tot}$ is defined above, \mbox{$t_0 = 0$} is the starting time of the simulation
\mbox{$t_{\rm c} = t_{\rm jet_1} + t_{\rm int}/2$} is the time exactly halfway between the end of the first jet
eruption and the beginning of the second jet eruption and the function $\tau_k(t)$ (with \mbox{$k=\{1,2\}$}) is
defined as:

\be
\tau_k(t) =\frac{1}{2}-\frac{1}{\pi} \times \arctan\left[\left(-1\right)^{k+1}
\frac{\zeta}{t_{\rm tot}} \times \left(  t - t_{\rm jet_k} \right) \right] \; .
\ee
\begin{comment}
The parameter $\zeta$ is a measure for the speed at function $\tau_k(t)$ `flips' around the point $t_{\rm jet_k}$.
\end{comment}
We find that \mbox{$\zeta=2744$} results in a fairly steep decline and rise of the jet inlet function near
\mbox{$t = t_{\rm jet_1}$} and \mbox{$t = t_{\rm jet_2}$}, while at the same time the temporal behavior of the
jets is well resolved.

\section{Results: stages in the evolution of a DDRG}
\label{sec:Results}

\subsection{Phase 1: the initial jet}

{\renewcommand{\arraystretch}{1.3}
\begin{table}
\caption{Jet-head advance speed \mbox{$\beta_{\rm hd}$} and effective impact radius \mbox{$R_{\rm am}$ [kpc]} for
the four different phases in episodic jet eruption. The results in phase 3 (the restarted jet
propagating entirely within the remnant cocoon) differ significantly from those in the phases 1, 2 and 4.
}
%\rowcolors{1}{white}{lightgray}
%\begin{minipage}{0.5\textwidth}
%\centering
  \begin{center}
%       {\bf Table 3}
%       \vskip \baselineskip
    \begin{tabular}{l c c c c} \hline
     \phantom{}                          & {\bf Phase}       & {\bf H2}   & {\bf I2}   & {\bf A2}
      \\ \hline
     $\beta_{\rm hd}$ ({jet 1})          & 1                 & 0.042      & 0.046      & 0.044
     \\
     \phantom{$\beta_{\rm hd}$} (jet 1)  & 2                 & 0.031      & 0.052      & 0.037
     \\
      \phantom{$\beta_{\rm hd}$} (jet 2) & 3                 & 0.703      & 0.732      & 0.701
     \\
     \phantom{$\beta_{\rm hd}$} (jet 2)  & 4                 & 0.040      & 0.038      & 0.036
     \\
     \phantom{}                          & \phantom{}        & \phantom{} & \phantom{} & \phantom{}
     \\
     $R_{\rm am}$ (jet 1)                & 1                 & 4.55       & 4.07       & 4.32
     \\
      \phantom{$R_{\rm am}$} (jet 1)     & 2                 & 6.11       & 3.58       & 5.20
     \\
     \phantom{$R_{\rm am}$} (jet 2)      & 3                 & 1.08       & 0.91       & 1.09
     \\
     \phantom{$R_{\rm am}$} (jet 2)      & 4                 & 4.78       & 4.99       & 5.33
     \\
     \\ \hline
    \end{tabular}

    \label{tab:EruptionPhase}
  \end{center}
%\end{minipage}
\end{table}
}

In phase 1 the first jet propagates through the $\uAM$, the same situation as treated in detail $\SWa$. 
This phase lasts approximately \mbox{15.3 Myr}. We will summarize the phase 1 results briefly, and refer
to $\SWa$ for more details.

\subsubsection{Jet-head advance speed}

All our jets are under-dense, with a proper density ratio between jet and surrounding medium roughly
\footnote{We take typical values since the isochoric (A) and isothermal (I) jet models have density
stratification.}
equal to \mbox{$\eta_{\rm R} = \rho_{\rm jt}/\rho_{\rm am} \sim 4.5 \times 10^{-3}$}. In such a situation
the jet-head advances into the ambient medium with a speed \mbox{$\beta_{\rm hd} \ll \beta_{\rm jt}$}, where
$\beta_{\rm hd}$ and $\beta_{\rm jt}$ are the jet-head advance speed and the bulk speed of the jet material
respectively in units of $c$ in the frame where the $\uAM$ is at rest. The jets develop significant structure
(long-lived vortices) that move with the jet-head, and therefore present an obstacle for the incoming $\uAM$, as
seen by an observer moving with the jet-head. As a result, the jet-head has an effective area
\mbox{$A_{\rm am} \equiv \pi \: R_{\rm am}^2$} perpendicular to the jet flow that is significantly larger than
the geometrical cross section \mbox{$A_{\rm jt} = \pi \: R_{\rm jt}^2$} of the undisturbed jet. Typically we
find that $A_{\rm am} \simeq (16-20) \times R_{\rm jt}$. The increased effective area determines the jet-head
advance speed. The relation between $\beta_{\rm hd}$, $\eta_{\rm R}$, $R_{\rm am}$ and $R_{\rm jt}$ is ($\SWa$,
Eqn. 58):
\be
\label{headadvance}
	\beta_{\rm hd} = \frac{\displaystyle \sqrt{\eta_{\rm R}} \:
        \gamma_{\rm jt} \beta_{\rm jt}}{\displaystyle \Omega + \sqrt{\eta_{\rm R}} \: \gamma_{\rm jt}}  \; ,
\ee
with $\Omega \equiv R_{\rm am}/R_{\rm jt}$.
\mbox{Figure \ref{fig:JHAS-2}} shows the jet-head advance speed for all three models, and covers the entire
simulation.  In phase 1 we find similar values for the advance speed between \mbox{10 Myr} and \mbox{15 Myr},
when the advance is more-or-less steady. Small differences between these simulations and those of $\SWa$ result
from our use of a different solver (HLLC instead of TVDLF).

By looking carefully at the simulations we find (as in $\SWa$) that the effective radius $R_{\rm am}$ also
roughly corresponds with the transverse size  of the {\em hotspots}, i.e. the region containing relativistically
hot gas that has gone through the {\em Mach disc}, the strong shock that effectively terminates the high-Mach
number jet flow.

{\renewcommand{\arraystretch}{1.3}
\begin{table}
\caption{This table shows the relativistic Mach numbers ($\mathcal{M}$) for both the Mach disc (MD) and the bow
shock (BS) at the jet-head for each of the three jet models $H2$, $I2$, $A2$, in phase 1 and phase 3. Moreover,
the ratio of the shock strength in phase 1 and phase 3 is given for the Mach disc, as well as for the bow shock.
}
%{\bf Table 4}
%\vskip \baselineskip
%\rowcolors{1}{white}{lightgray}
%\begin{minipage}{0.5\textwidth}
%\centering
\begin{center}
    \begin{tabular}{l c c c c} \hline
     \phantom{}                                & {\bf Phase} & {\bf H2}   & {\bf I2}   & {\bf A2}
      \\ \hline
     {\em $\mathcal{M}_{\rm MD_1}$}         & 1           & 13.3       & 21.7       & 13.2
     \\
     {\em $\mathcal{M}_{\rm MD_2}$}         & 3           & 10.9       & 15.7       & 7.58
     \\
     {\em $\mathcal{M}_{\rm MD_1}/\mathcal{M}_{\rm MD_2}$}
                                               & \phantom{}  & 1.22       & 1.38       & 1.74
     \\
     \phantom{}                                & \phantom{}  & \phantom{} & \phantom{} & \phantom{}
     \\
     {\em $\mathcal{M}_{\rm BS_1}$}         & 1           & 32.6       & 35.7       & 34.1
     \\
     {\em $\mathcal{M}_{\rm BS_2}$}         & 3           & 2.21       & 2.40       & 2.20
     \\
     {\em $\mathcal{M}_{\rm BS_1}/\mathcal{M}_{\rm BS_2}$}
                                               & \phantom{}  & 14.7       & 14.8       & 15.5
     \\ \hline
    \end{tabular}

    \label{tab:MachNumber}
\end{center}
%\end{minipage}
\end{table}
}

\subsubsection{Temporal behavior along the jet axis}
\label{subsubsec:TemporalBehavior}

\begin{figure*}
%\begin{center}
$
\begin{array}{c c}
\includegraphics[clip=false,trim=0cm 0cm 0cm 0cm,width=0.45\textwidth]{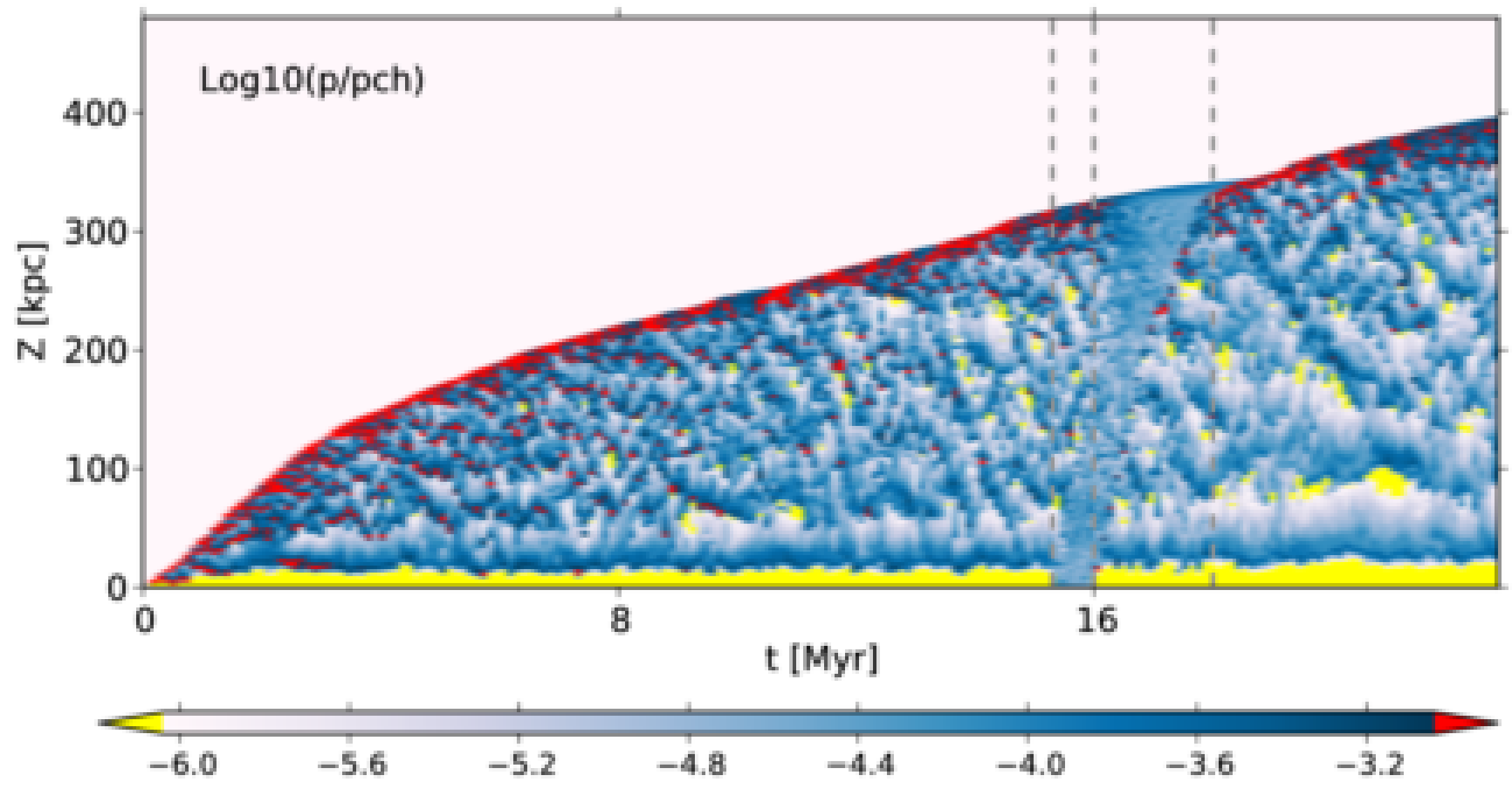}&
\includegraphics[clip=false,trim=0cm 0cm 0cm 0cm,width=0.45\textwidth]{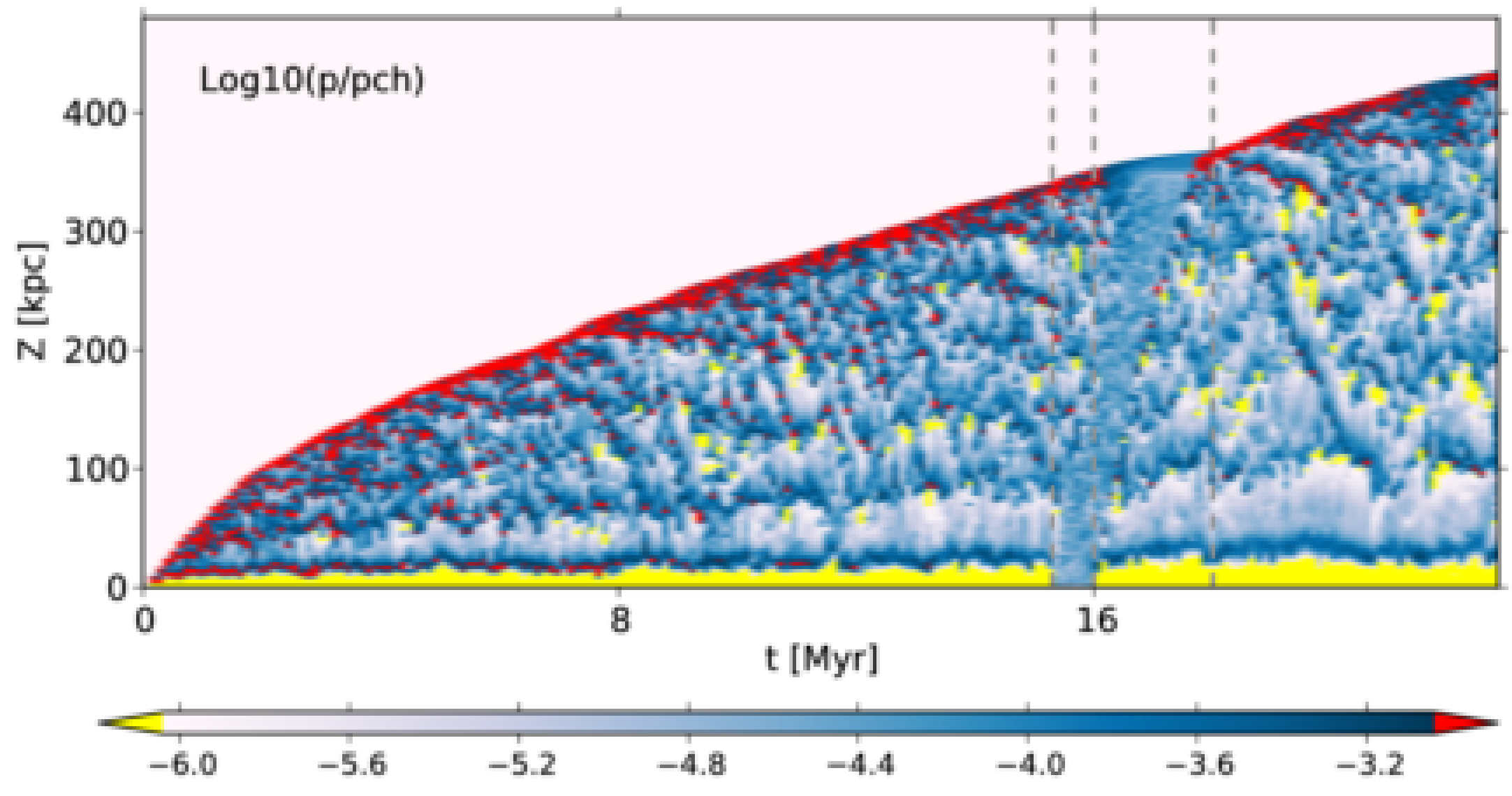} \\
\includegraphics[clip=false,trim=0cm 0cm 0cm 0cm,width=0.45\textwidth]{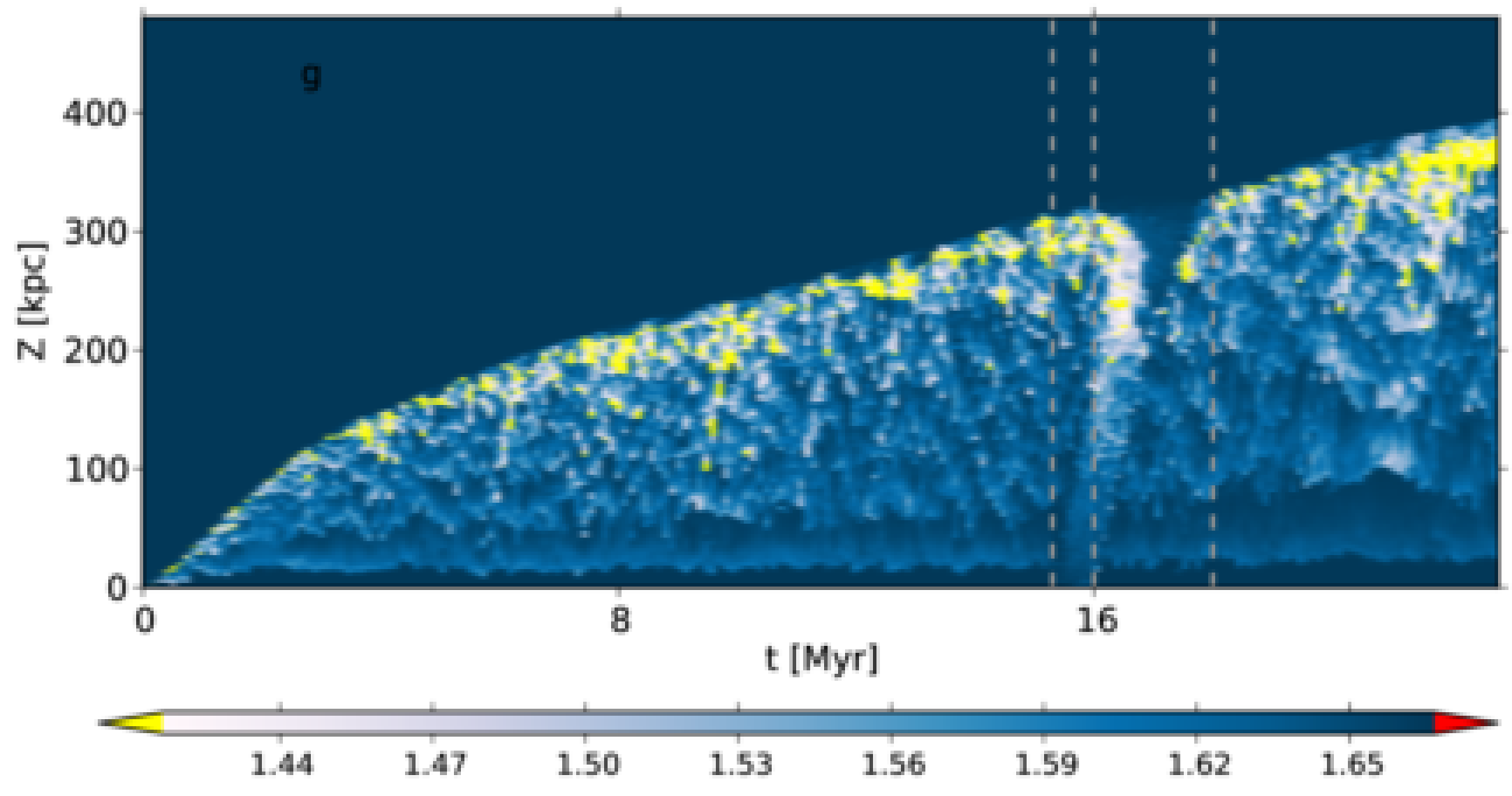}&
\includegraphics[clip=false,trim=0cm 0cm 0cm 0cm,width=0.45\textwidth]{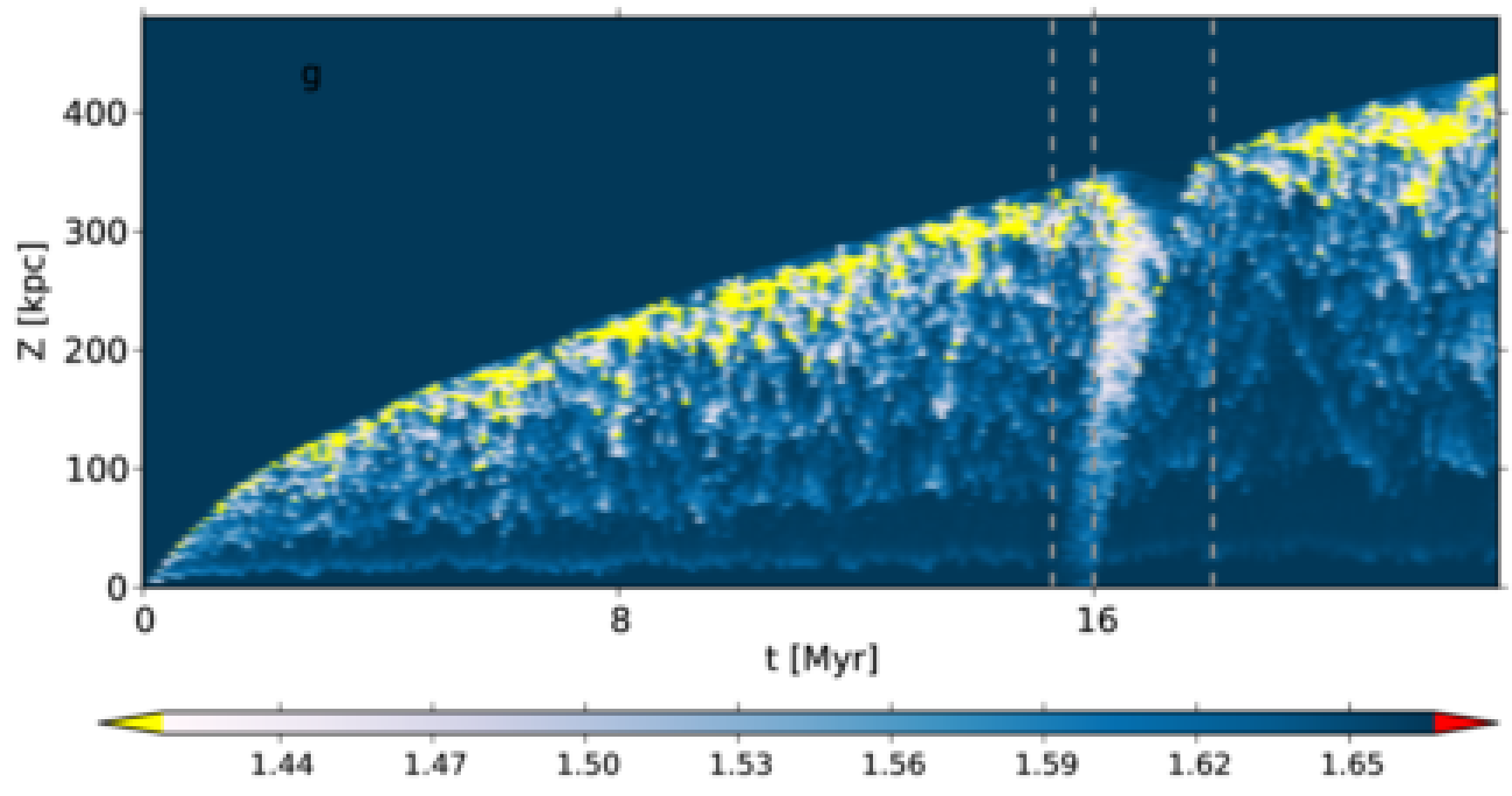} \\
\includegraphics[clip=false,trim=0cm 0cm 0cm 0cm,width=0.45\textwidth]{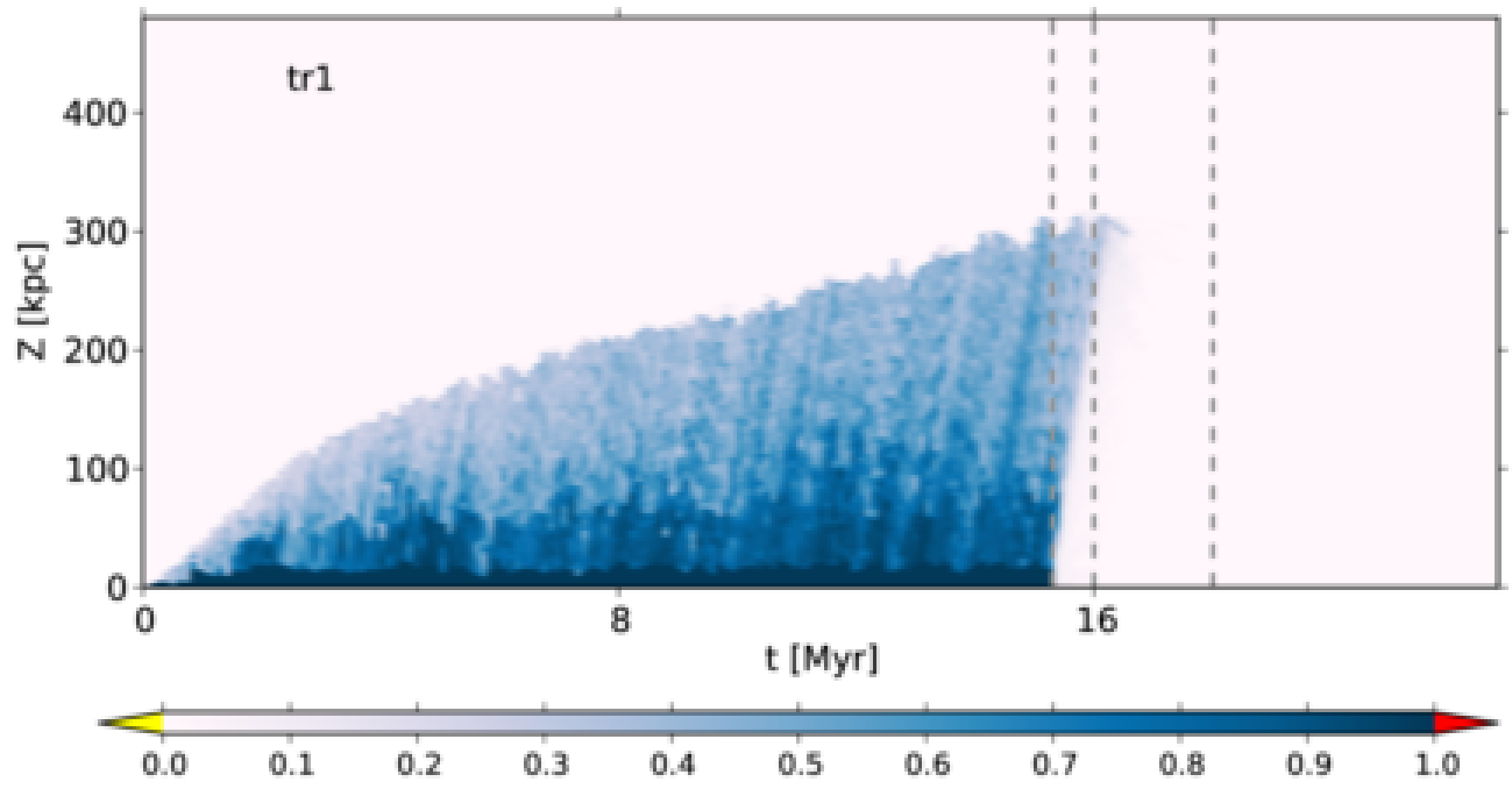}&
\includegraphics[clip=false,trim=0cm 0cm 0cm 0cm,width=0.45\textwidth]{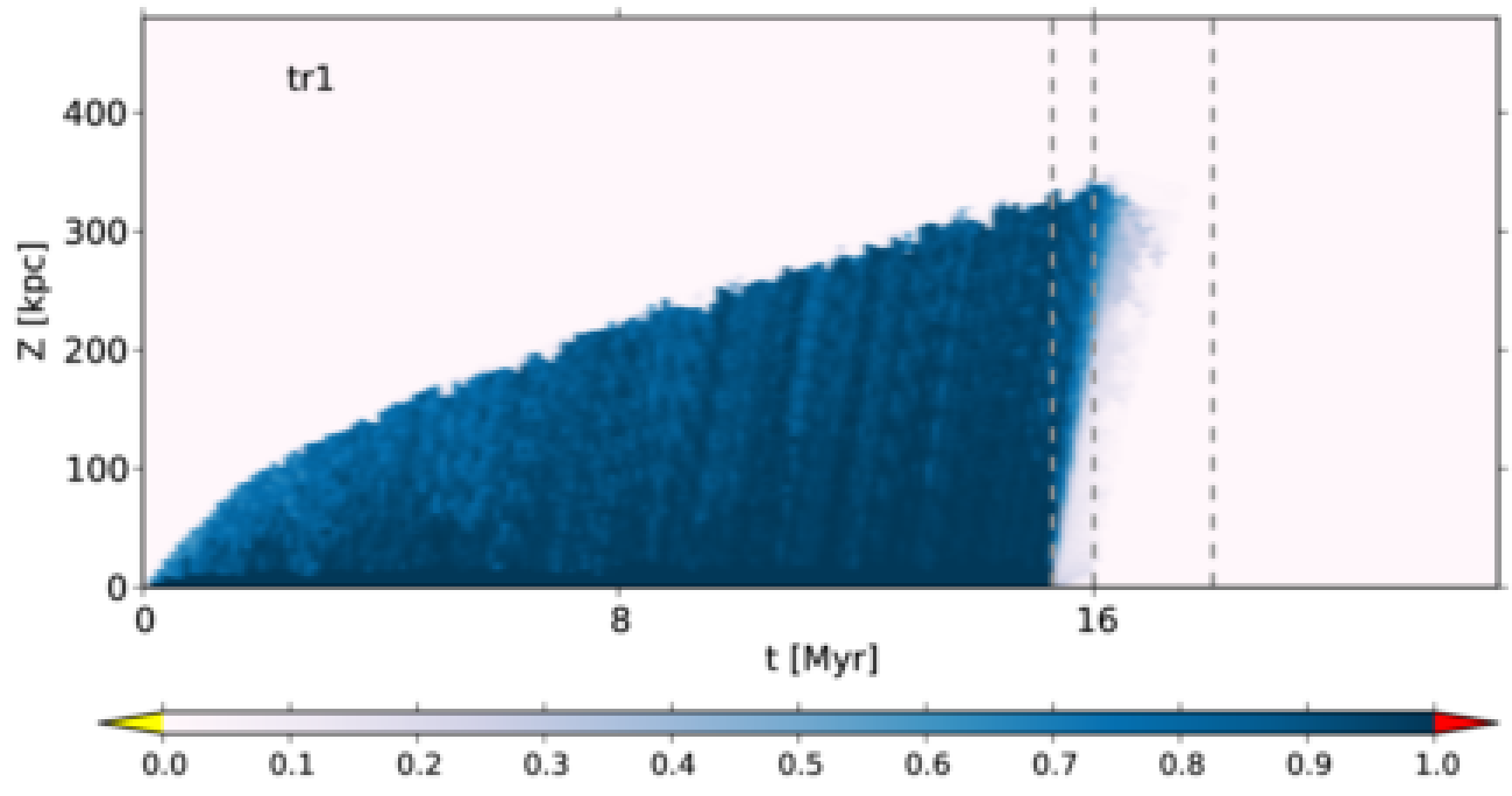} \\
\includegraphics[clip=false,trim=0cm 0cm 0cm 0cm,width=0.45\textwidth]{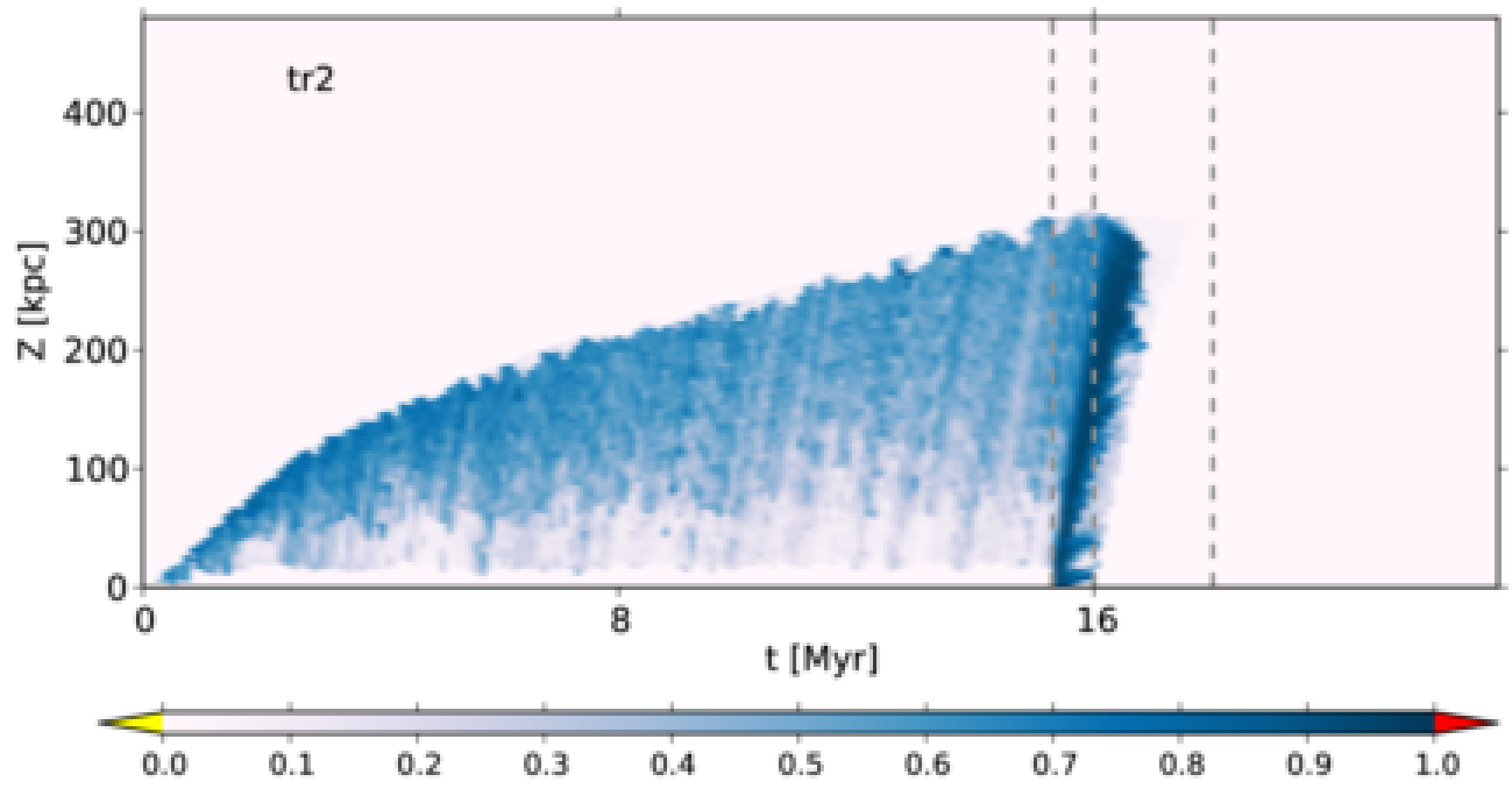}&
\includegraphics[clip=false,trim=0cm 0cm 0cm 0cm,width=0.45\textwidth]{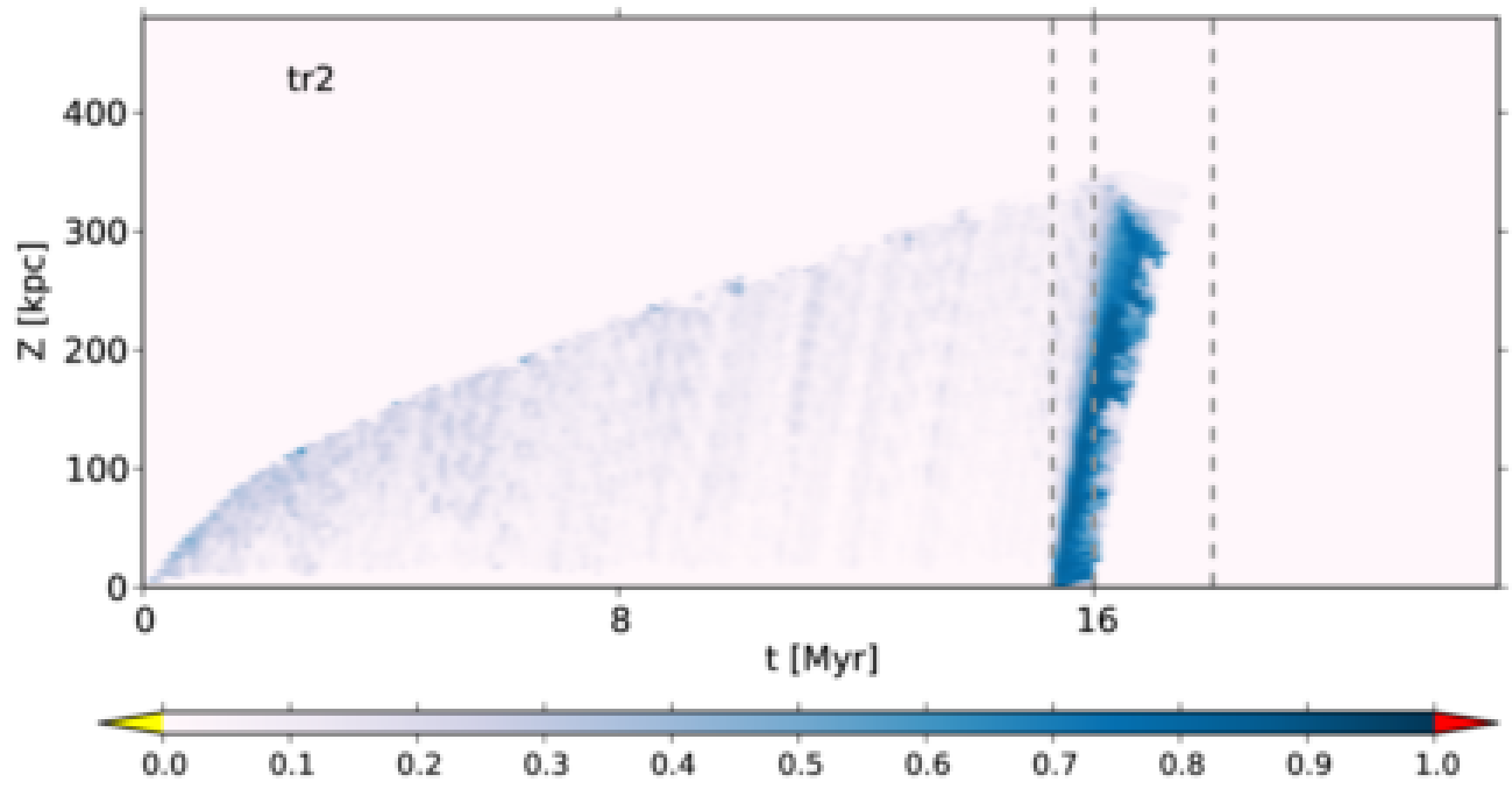}
\end{array}
$
%\end{center}
%\vspace{1.5 cm}
\caption{Temporal behavior of different variables as measured along the jet axis. The left column shows the time
plots of the isochoric jet ($A2$) and the right column show the time plots of the isothermal jet ($I2$). From top
to bottom the following variables are shown: gas pressure ($log(p/{p_{\rm ch}}$); effective polytropic index
($\Gamma_{\rm eff}$); tracer value of the jet spine of the first jet ($\theta_1^{\rm sp}$) and tracer value of
the jet sheath of the first jet ($\theta_1^{\rm sh}$). The three vertical dashed lines separate the four
different phases of the episodic event. The regions in red exceed the upper threshold, while the regions in
yellow drop below the lower threshold. In the plot of the pressure, the red region shows where the gas
pressure exceeds a threshold of \mbox{$P = 10^3 P_0$}, with $P_0$ the gas pressure of the jet at the jet inlet.
This high-pressure region coincides with the hotspot of the jet.
}
  \label{fig:TimePlots}

\end{figure*}

\mbox{Figure \ref{fig:TimePlots}} shows a cut along the $Z-$axis of a number of hydrodynamical quantities as
a function of time. In these plots, phase 1 of the episodic jet event is contained between the left boundary of
the panels and the first dashed line at \mbox{$t = 15.3$ Myr}. The left column shows the plots for the isochoric
jet $A2$ and the right column shows the plots for the isothermal jet $I2$. The pressure panels (top row) show
an adjustment shock close to the jet inlet with an approximate constant distance to the jet inlet. Directly after
the adjustment shock, the pressure in the jet increases significantly and shows large fluctuations (internal
shocks) along the jet axis. At the jet-head (corresponding to the inclining line at the top of the disturbed
regions), the hotspots can be recognized by the strong increase in pressure, as denoted by the red dots. The
second row (effective polytropic index $\Gamma_{\rm eff}$) shows similar behavior: near the jet inlet the jet
plasma is non-relativistically cold with \mbox{$\Gamma_{\rm eff} \approx 5/3$}. Nearing the jet-head, the gas
becomes hotter and at the jet-head the gas is shocked to (near-)relativistic temperatures with
\mbox{$\Gamma_{\rm eff} \lesssim 1.417$}. Finally, the third row and the bottom row show the tracer values
of jet-spine and jet-sheath of the first jet. The difference in transverse structural integrity is well
reflected in these plots: for the isochoric jet, immediately after the adjustment shock jet-spine material mixes
with material of the jet-sheath and transverse structure is lost. However, in case of the isothermal jet, the
jet-spine tracer abundance is nearly unaffected as material flows towards the jet-head.

%\begin{comment}
%%% Time plots of log P etc. along the Z-axis
%\end{comment}

\subsubsection{Cocoon structure and mixing}

The large pressure near the jet-head and the low jet-head advance speed (\mbox{$\beta_{\rm hd_1} \simeq 0.04$})
imply that a thick cocoon is formed with a significant back-flow. In this cocoon, shocked jet material (gas that
has gone through the Mach disc) and shocked intergalactic gas (that has gone through the bow shock in the $\uAM$
that precedes the jet-head) mix efficiently. This mixing is facilitated by the vortices (and the ensuing
turbulence) that are shed by the jet-head at quasi-regular intervals. These vortices then lead to pressure
fluctuations in the cocoon that, when transmitted to the jet flow, lead to the formation of internal shocks in
the jet.

In the case of the homogeneous jet (model $H2$) these internal shocks have only a small influence on jet
structure: the jet transverse structure is almost completely maintained until it reaches the Mach disc. For the
isochoric jet $A2$ the density jump between spine and sheath leads to shock reflection. This creates an internal
flow where efficient mixing of spine and sheath material occurs, as illustrated by the jet spine/sheath tracer
values in the top-left panel of \mbox{Figure \ref{fig:ZcutsA2}}, and the left panel of
\mbox{Figure \ref{fig:RcutsA2}}. In the isothermal jet (model I2) the density jump between spine and sheath is
absent, and the amount of spine--sheath mixing is reduced significantly compared to model A2.

\subsection{Phase 2: from jet switch-off to jet restart}

In the second phase, the initial jet is no longer driven by the AGN, so that no fresh jet material enters the
system. However, at the moment that the inflow stops, the fast-flowing jet material that is still present
continues to move toward the jet-head (driven by inertia). This continues until all the material has gone through
the Mach disc. As long as this flow is still present, the bulk of that jet material maintains Lorentz factors of a
few and jet material approaches the jet-head with a velocity close to $c$. 

Eventually, the trailing end of the initial jet will reach the Mach disc, after which the initial jet flow will
have completely disappeared. Presumably, the production of relativistic electrons at shocks (Fermi-I
acceleration) will quickly cease, while any acceleration by turbulence (Fermi-II acceleration) will gradually
become less and less important as turbulence dissipates in the hotspots and radio lobes. Therefore (depending on
frequency and cooling time of the non-thermal electrons), the hotspots and radio lobes should start to dim. The
time it takes for the trailing end of the jet to be terminated at the Mach disc once the jet has been switched
off is therefore determined by the length of the cocoon, $D_{\rm co}$. In all three models we have
\mbox{$D_{\rm co} \simeq 350$ kpc}. With \mbox{$\beta_{\rm jt} \simeq 0.95$} (a typical bulk velocity along the
jet axis) one expects that it  takes approximately \mbox{$D_{\rm co}/c \beta_{\rm jt} \simeq 1.1$ Myr} for the
hotspots to turn off and for the first jet to disappear. This value is in close agreement with the results of our
simulations.

The average jet-head advance speed for the jets in phase 2 is calculated between \mbox{$t = 15.3$ Myr} and
\mbox{$t = 16.3$ Myr} and shown in \mbox{Table \ref{tab:EruptionPhase}}. It shows that the jet-heads continue to
propagate with approximately the same velocity as in phase 1 as long as the initial jets have not completely
disappeared. This results in an effective impact area $A_{\rm am}$ comparable to that found in phase 1.

We note that in case of the isothermal and the isochoric jet, the jet spine has a higher Lorentz factor than the
jet sheath.
Therefore, the trailing end of the jet spine outruns the trailing end of the jet sheath. This initially creates a
cavity along the jet axis at the trailing end of the jet spine, locally resulting in a strong radial pressure
gradient. As a result, the material from the direct surrounding of the cavity (which at that point is
the jet sheath) flows towards the $z$-axis where it fills up the cavity almost instantly. In the process of
filling up the cavity, the forward momentum of that material is lost, leaving behind patches of material that
originate from the jet sheath. At the trailing end of the jet sheath, a similar process occurs, now involving
cocoon material: again a cavity is formed with the associated pressure gradients. This cavity gets filled up by
material from its immediate surroundings, in this case mostly cocoon material. Therefore, behind the trailing end
of the jet one predominantly finds cocoon material and patches that contain higher concentrations of material
originating from the jet sheath along the old jet path.
Many of these
events can be recognized in \mbox{Figure \ref{fig:TimePlots}}. Here, phase 2 extends from the left dashed line
(at \mbox{$t = 15.3$ Myr}) to the dashed line in the middle (at \mbox{$t = 16.0$ Myr}). As soon as the first jet
is switched off, the adjustment shock near the jet inlet disappears. The front end of the jet continues to
propagate toward the Mach disc and then disappears around \mbox{$t \approx 16.6$ Myr}. A bar-shaped
feature can be recognized in the panels for the pressure ($\log(P/P_{\rm ch})$), effective polytropic index
($\Gamma_{\rm eff}$) and tracer of the jet-sheath ($\theta_{\rm 1}^{\rm sh}$). This feature roughly stretches
from \mbox{$\sim 15.3 - 16$ Myr} at the bottom of the panels up to \mbox{$\sim 16 - 18$ Myr} at the top. It
corresponds to the region behind the trailing end of the switched-off jet. In this bar-shaped structure, a strong
increase in concentration of the jet-sheath is found for both the isochoric and the isothermal jet. It is a
result of the void-filling jet-sheath material, due to the escaping jet-spine along the jet axis.

\subsection{Phase 3: propagation of the second jet in the remnant cocoon}
\label{subsec:Phase3}

%\begin{comment}
%%% Cuts along the $Z-$ axis of the $A2$, line plots:
\begin{figure*}
%\begin{center}
$
\begin{array}{c c}
\includegraphics[clip=true,trim=2.5cm 2.5cm 1cm 1cm,width=0.5\textwidth]{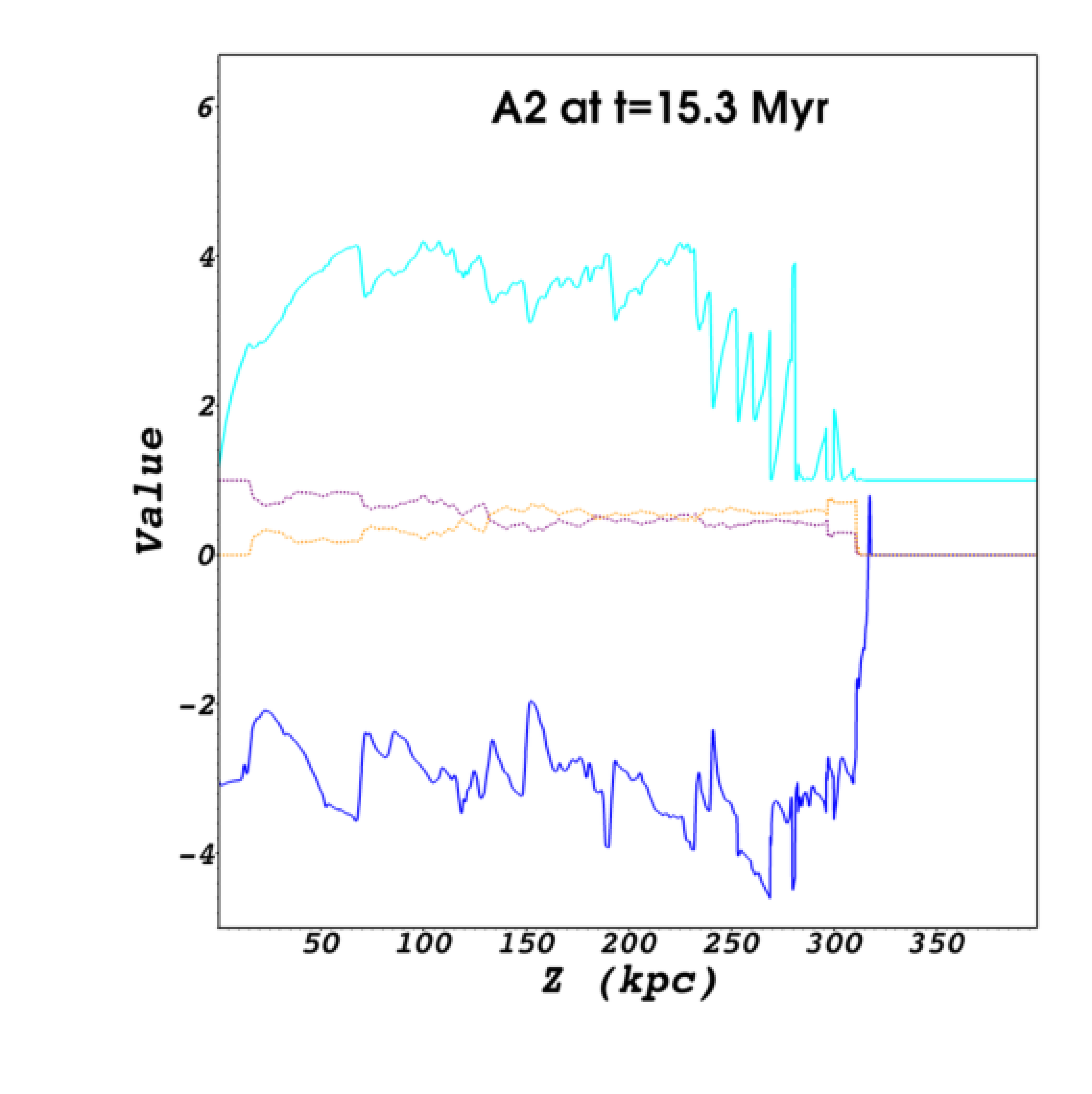}&
\includegraphics[clip=true,trim=2.5cm 2.5cm 1cm 1cm,width=0.5\textwidth]{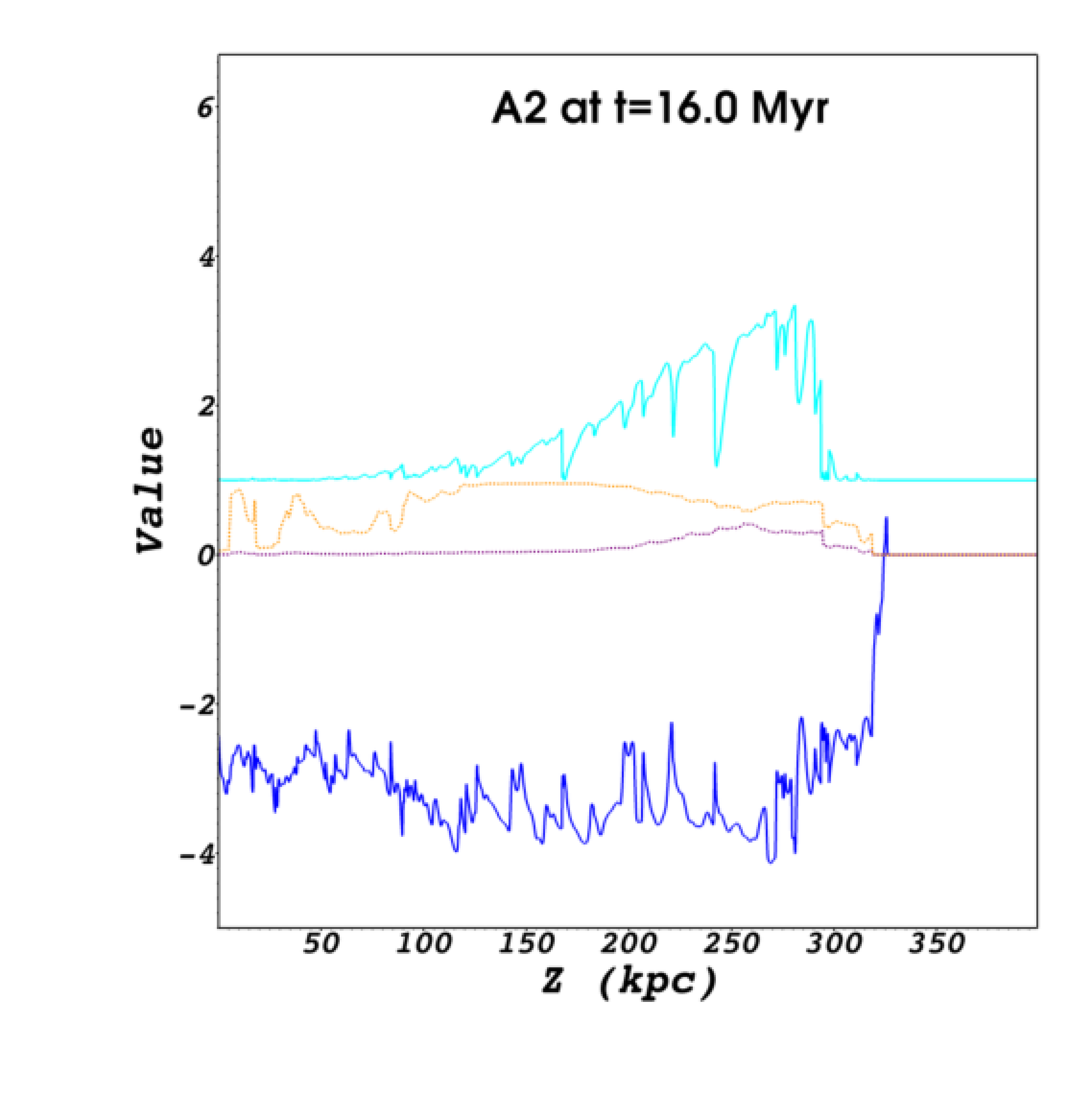}\\
\includegraphics[clip=true,trim=2.5cm 2.5cm 1cm 1cm,width=0.5\textwidth]{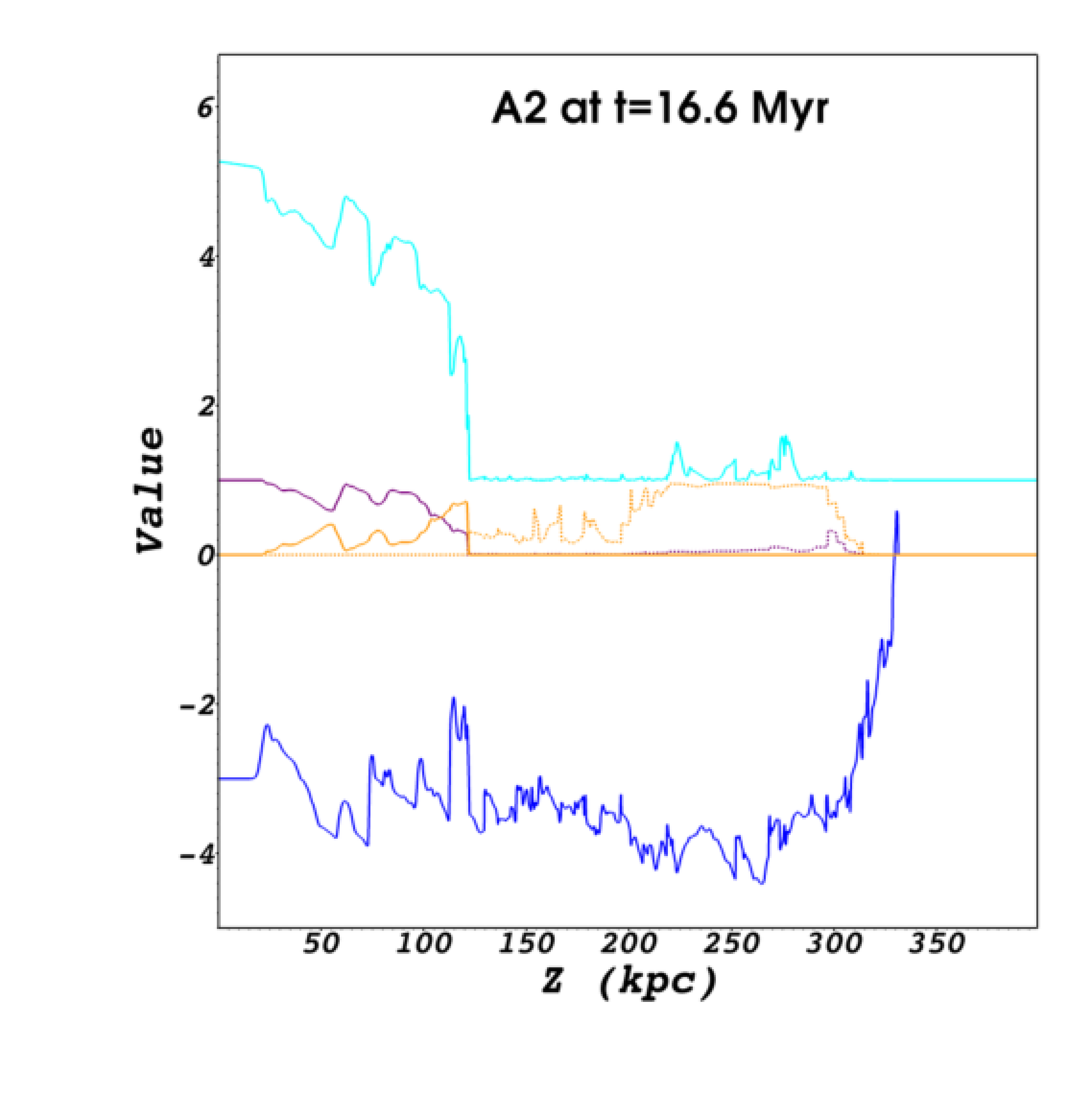}&
\includegraphics[clip=true,trim=2.5cm 2.5cm 1cm 1cm,width=0.5\textwidth]{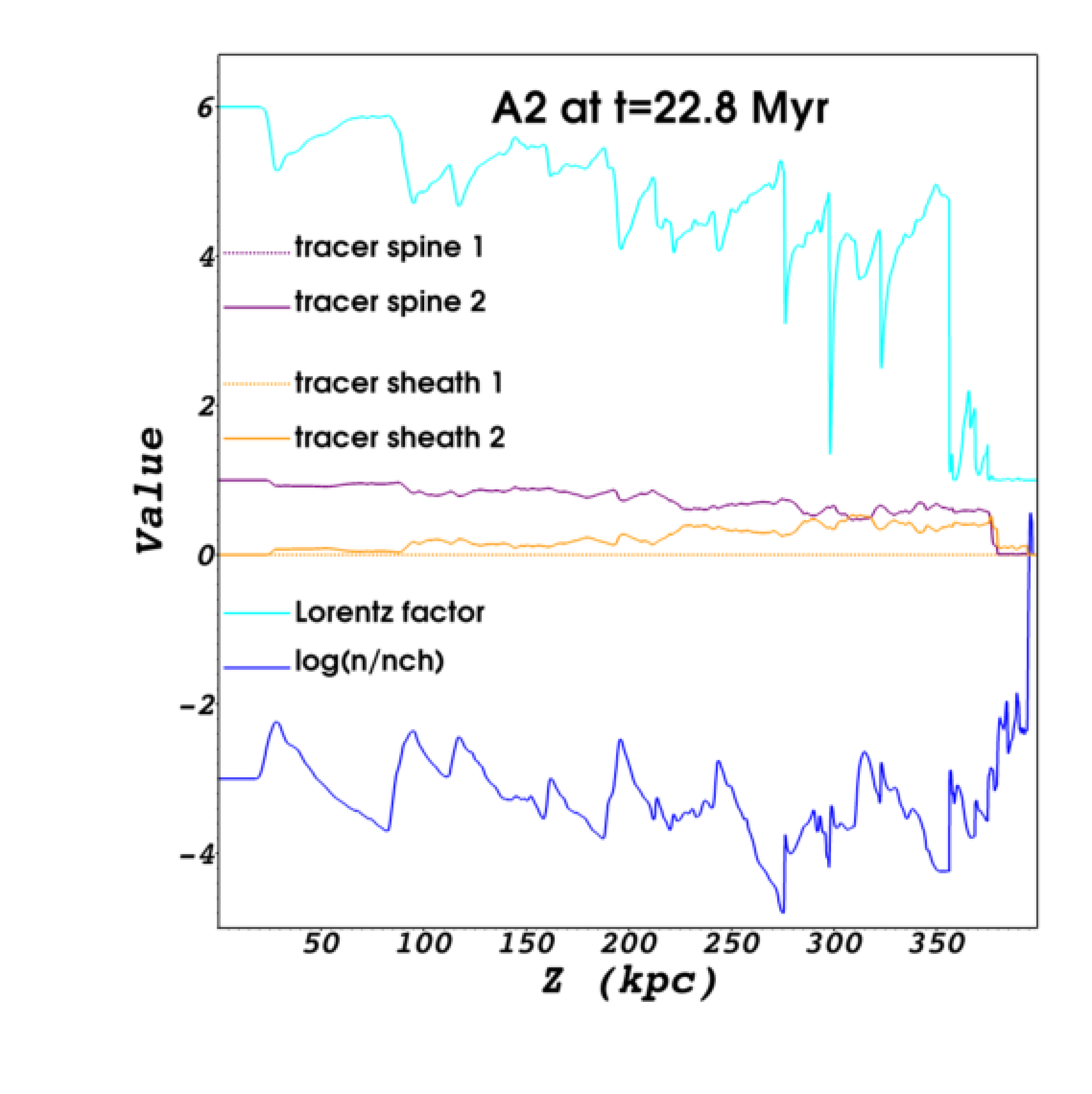}
\end{array}
$
%\end{center}
%\vspace{1.5 cm}
\caption{Cuts along the jet ($Z$-) axis of the isochoric jet $A2$ at a characteristic time of each of the four
different phases in episodic jet activity. In the top left plot, the initial jet has been injected into the
system for a total time of \mbox{$t$ = 15.3 Myr} and is about to be switched off. It marks the end of the first
phase and the beginning of the second phase. In the top right plot at \mbox{$t$ = 16.0 Myr}, the initial jet has
been switched off for \mbox{$\sim$ 0.7 Myr} and the restarting jet is about to be injected into the system.
Therefore, this plot marks the end of phase 2 and the beginning of phase 3. In the bottom left plot at
\mbox{$t$ = 16.6 Myr}, the restarted jet is propagating completely within the $\dAM$ that was left by the
initial jet. At this time, for all three jet models the initial jet has almost completely disappeared.
Therefore, this time frame marks an overlap between phase 2 and phase 3. In the bottom right plot at
\mbox{$t$ = 22.8 Myr}, the jet has penetrated the forward edge of the $\dAM$ and is therefore propagating
in the $\uAM$. This marks phase 4. The curves in dark blue show the number density $\log(n/n_{\rm ch})$. The
curves in light blue shows the bulk Lorentz factor $\gamma$. Finally, the curves in purple and orange show the
tracer values of the jet spine $\theta_{\rm i}^{\rm sp}$ and jet sheath $\theta_{\rm i}^{\rm sh}$ respectively.
Here, the dashed lines stand for the initial jet \mbox{($i = 1$)}, while the solid lines stands for the
restarted jet \mbox{($i = 2$)}.
}
  \label{fig:ZcutsA2}

\vspace{10 cm}
\end{figure*}
%\end{comment}

%\begin{comment}
%%% Cuts along the $R-$ axis at Z = 90 kpc of the $A2$, line plots:
\begin{figure*}
%\begin{center}
$
\begin{array}{c c}
\includegraphics[clip=true,trim=2.5cm 2.5cm 1cm 1cm,width=0.5\textwidth]{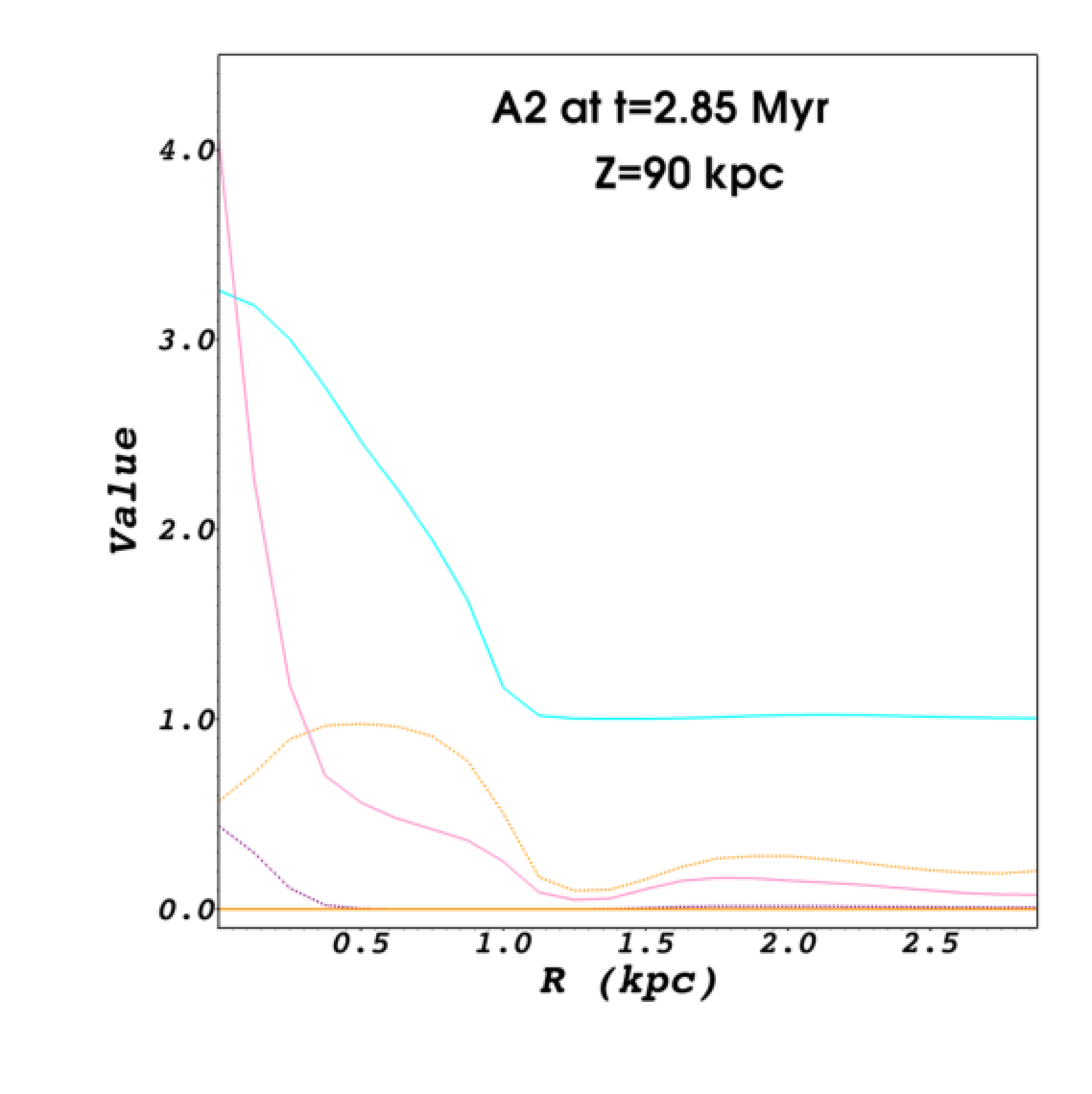}&
\includegraphics[clip=true,trim=2.5cm 2.5cm 1cm 1cm,width=0.5\textwidth]{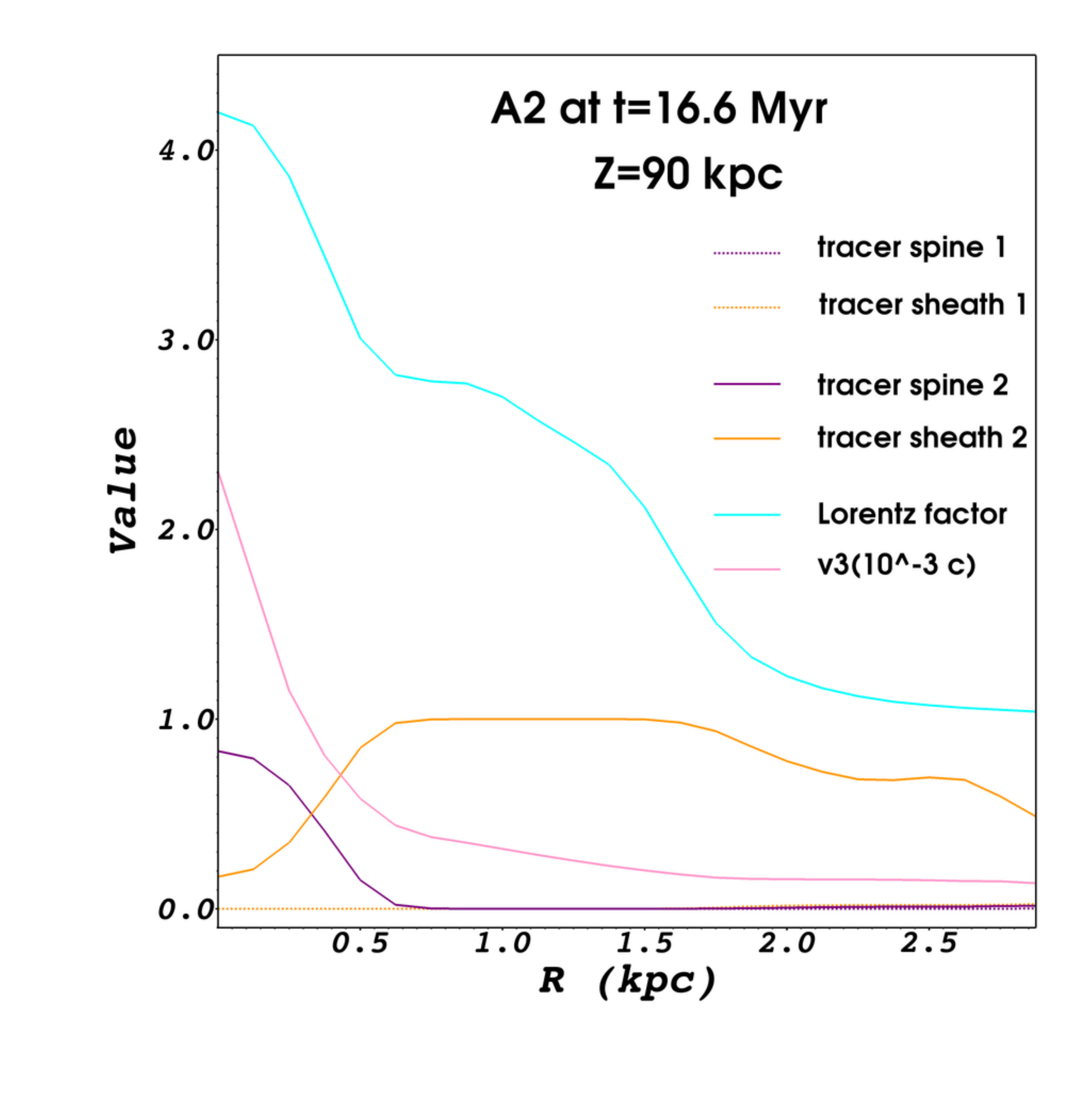}
\end{array}
$
%\end{center}
\caption{Radial cuts along the jet cross section of the isochoric jet $A2$ at a height of $Z = $ 90 kpc. The plot
on the lefthand side shows the first jet (in phase 1) at \mbox{$t = 2.85$ Myr}, whereas the plot on the righthand
side shows the second jet (in phase 3) at \mbox{$t = 16.6$ Myr}. In order to make a fair comparison, the time
frames are chosen such that the length of the jets, \mbox{$D_{\rm co}$}, are approximately equal. In this case we
choose \mbox{$D_{\rm co} \approx 121$ kpc}. The colors of the curves are similar to those in
\mbox{Figure \ref{fig:ZcutsA2}}. In addition, the curve in pink shows the azimuthal velocity $v_{\phi}$ in units
of \mbox{$10^{-3}$ c}. The transverse structural integrity at \mbox{$Z = 90$ kpc} of the second jet (compared to
the structural integrity at jet inlet) is stronger maintained than for the first jet. This behavior is seen for
most heights $Z$ along the jet axis.
}
  \label{fig:RcutsA2}

%\vspace{10 cm}
\end{figure*}
%\end{comment}

During phase 3, the central engine goes through a period of renewed activity, somehow restarting the jet.
We have chosen to start the second jet with the same intrinsic parameters (such as diameter, structure,
luminosity, pressure profiles, etc.) as the first jet in order to make a simple comparison between the properties
of the two jets possible. Once the second jet has started, it propagates within the cocoon left by the first jet
($\dAM$).
\footnote{Although the first jet is set up in pressure equilibrium with its ambient medium, the second jet is
injected into a medium that is over-pressured by a factor of 10-100. However, in practice this does not lead to
a strong difference in the propagation and evolution of the second jet near the jet inlet. The reason is as
follows: as soon as the first jet is injected into the uIGM, the strong shocks at the jet-head shock-heat the
gas, creating the hot and inflated cocoon almost instantaneously with a pressure 10-100 times that of the jet
itself. As a result, the jet is compressed to re-establish pressure balance, leading to a re-adjustment shock
inside the jet, close to the jet inlet. When the second jet is injected into the dIGM, it also
encounters an over-pressured ambient medium, resulting in the formation of a similar adjustment shock at
approximately the same distance to the jet inlet (this can for example also be seen in the upper panels of
\mbox{Figure \ref{fig:TimePlots}}).}

The $\dAM$ shows significant turbulence, with large fluctuations in mass density and  pressure. The mass
density of the $\dAM$ (\mbox{$\rho_{\rm am} = \rho_{\rm dIGM}$}) is now a factor \mbox{$\sim 10^2-10^4$}
smaller than in the $\uAM$ (\mbox{$\rho_{\rm am} = \rho_{\rm uIGM}$}), and is of the same order as the mass
density in the jets. Defining as before the density ratio $\eta_{\rm R} \equiv \rho_{\rm jt}/\rho_{\rm am}$ this
parameter now takes a typical value \mbox{$0.45 \lesssim \eta_{\rm R} \lesssim 45$}, close to unity.
The gas pressure in the $\dAM$ is typically a factor \mbox{$10-100$} higher than the pressure of the $\uAM$ due
to the shock-heated gas that resides in the $\dAM$, see for example the top left panel of
\mbox{Figure \ref{fig:I2Phase34}}.

\subsubsection{Jet-head advance speed and the strength of Mach disc and bow shock}

The change in $\eta_{\rm R}$, from $\eta_{\rm R} \sim 4.5 \times 10^{-3}$ for the (under-dense) first jet to
$\eta_{\rm R} \sim 1$ for the second jet, leads to a much higher jet-head advance speed. This is easily seen in
\mbox{Figure \ref{fig:JHAS-2}}. This means that the velocity with which the jet material enters the Mach disc
drops, as does the strength of the shock (i.e. the proper Mach number $\mathcal{M}_{\rm MD}$ of the Mach disc,
see below). 

The increase of $\beta_{\rm hd}$ (from \mbox{$\beta_{\rm hd_1} \approx 0.04$} to
\mbox{$\beta_{\rm hd_2} \approx 0.7$}, see \mbox{Table \ref{tab:EruptionPhase}}) increases the velocity with
which material enters the bow shock of the second jet. However, the effect of the increased velocity is more
than offset by the effect of the large temperature increase accompanying the increased pressure and decreased
density in the $\dAM$. This temperature increase leads to an increase by a factor of \mbox{$\sim 400$} in the
sound speed, $c_{\rm s}$, in front of the bow shock
\footnote{We find \mbox{$c_{\rm s}($\uAM$) \approx 1.3 \times 10^{-3} \: c$} and
\mbox{$c_{\rm s}($\dAM$) \approx 0.41 \: c$}}.
As a result, the strength of the bow shock (Mach number $\mathcal{M}_{\rm BS}$) {\em also} decreases: the bow
shock preceding the head of the second jet is not as strong as the bow shock preceding the first jet. The
conclusion is that the two shocks associated with the jet-head of the second jet are both weaker than those
associated with the first jet.

We have calculated the relativistic Mach numbers for the Mach disc and the bow shock at the jet-head
for each of the three jet models $H2$, $I2$ and $A2$ in phase 1 and phase 3. The results are given in
\mbox{Table \ref{tab:MachNumber}}. The quantity determining shock strength (and shown in
\mbox{Table \ref{tab:MachNumber}}) is the {\em proper relativistic Mach number} \citep{Konigl1980}: 
\be
\mathcal{M} \equiv \frac{u_{\rm sh}}{u_{\rm s}} =
            \frac{\gamma_{\rm sh}\beta_{\rm sh}}{\gamma_{\rm c_{\rm s}}c_{\rm s}} \; .
\ee
Here $\beta_{\rm sh}$ is the velocity of the incoming flow along the shock normal and
\mbox{$\gamma_{\rm sh} = 1/\sqrt{1 - \beta_{\rm sh}^2}$} the associated Lorentz factor. 
The sound velocity of the gas is given by:
\be
\beta_{\rm s} = \frac{c_{\rm s}}{c} = \sqrt{\frac{\Gamma_{\rm eff} P}{\rho h}} \; ,
\ee
with \mbox{$h = 1 + (e_{\rm th} + P)/\rho$} the specific relativistic enthalpy (as follows from Eqn. 7 and 8
in \citealt{Keppens2012}). Also \mbox{$\gamma_{\rm s} = 1/\sqrt{1 - \beta_{\rm s}^2}$}.
\nskip
The jet material enters the Mach disc with velocity $\beta_{\rm sh} \equiv \beta_{\rm MD}$ equal to:
\be
\beta_{\rm MD} = \frac{\beta_{\rm jt} - \beta_{\rm hd}}{1 - \beta_{\rm jt}\beta_{\rm hd}} \; .
\label{eq:betaRel}
\ee
The corresponding Lorentz factor is:
\be
\gamma_{\rm MD} = \gamma_{\rm jt}\gamma_{\rm hd}(1 - \beta_{\rm jt}\beta_{\rm hd}) \; .
\label{eq:gammaRel}
\ee 
The bow shock advances into the $\dAM$ with speed $\beta_{\rm sh} \simeq \beta_{\rm hd}$, neglecting the small
lab-frame velocity of the $\dAM$ material itself.

\mbox{Tables \ref{tab:EruptionPhase} and \ref{tab:MachNumber}} summarize and compare the results for jet 1 in
phases 1 and 2, and jet 2 in phases 3 and 4. As can be seen in Figure \ref{fig:JHAS-2}, the restarted jets
advance through the remnant cocoon with almost equal velocities: $\beta_{\rm hd_2} \simeq 0.7$. We calculated the
average jet-head advance speed for the restarted jets between \mbox{$t = 16.0$ Myr} and \mbox{$t = 17.0$ Myr}
(see \mbox{Table \ref{tab:EruptionPhase}}). In that time interval, the restarted jets have not yet broken out of
the remnant cocoon. It is immediately seen that the jet-head advance speed of the restarted jets is much larger
than that of the initial jets by a factor of \mbox{$\sim$ 16}. The reason for this significant change in advance
speed is twofold: [1] the fact that the mass density ratio between the $\dAM$ and the jet is now
\mbox{$\eta_{\rm R} = \rho_{\rm jt} / \rho_{\rm am} \approx 1$} and [2] the fact that the effective impact area
of the restarted jet is much reduced, close to its geometrical cross section:
\mbox{$\Omega = R_{\rm am}/R_{\rm jt}  \simeq 1$}. According to relation (\ref{headadvance}) both effects lead to
an increase of $\beta_{\rm hd}$.

\subsubsection{Mass discharge and cocoon size} 

\begin{figure*}
%\begin{center}
$
\begin{array}{c c}
\includegraphics[clip=true,trim=2.5cm 2.5cm 1cm 1cm,width=0.5\textwidth]{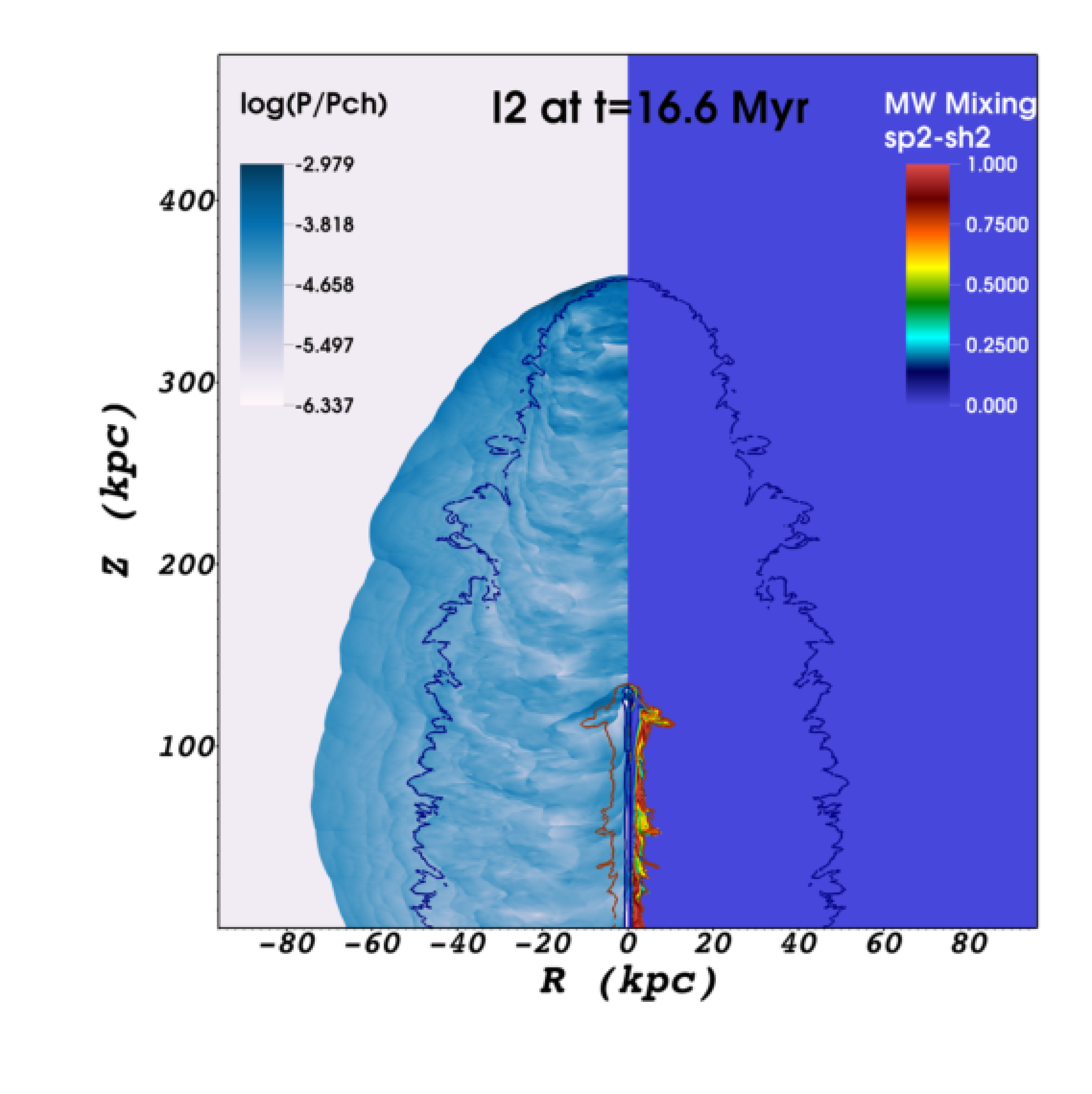}&
\includegraphics[clip=true,trim=2.5cm 2.5cm 1cm 1cm,width=0.5\textwidth]{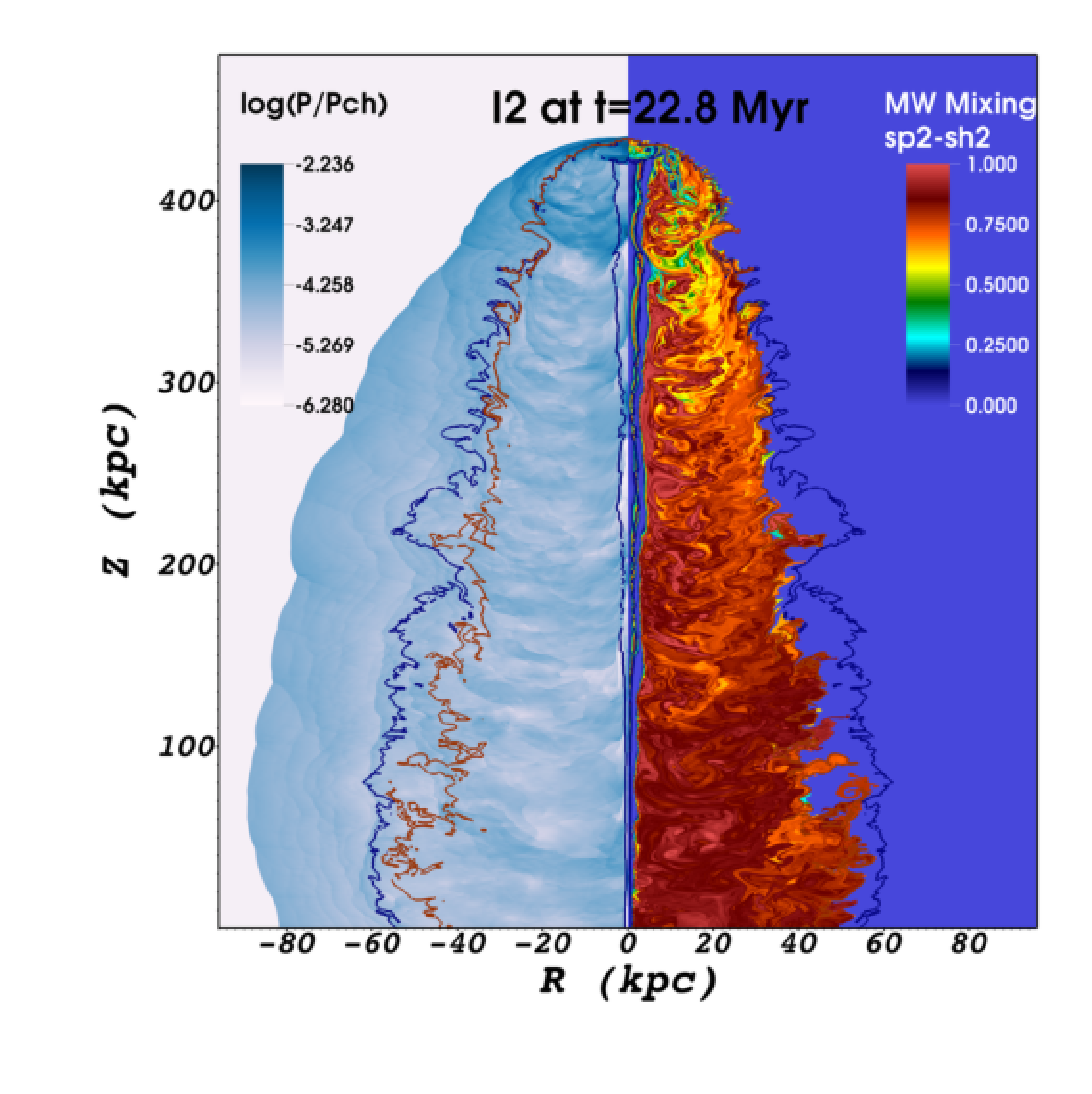}\\
\includegraphics[clip=true,trim=2.5cm 2.5cm 1cm 1cm,width=0.5\textwidth]{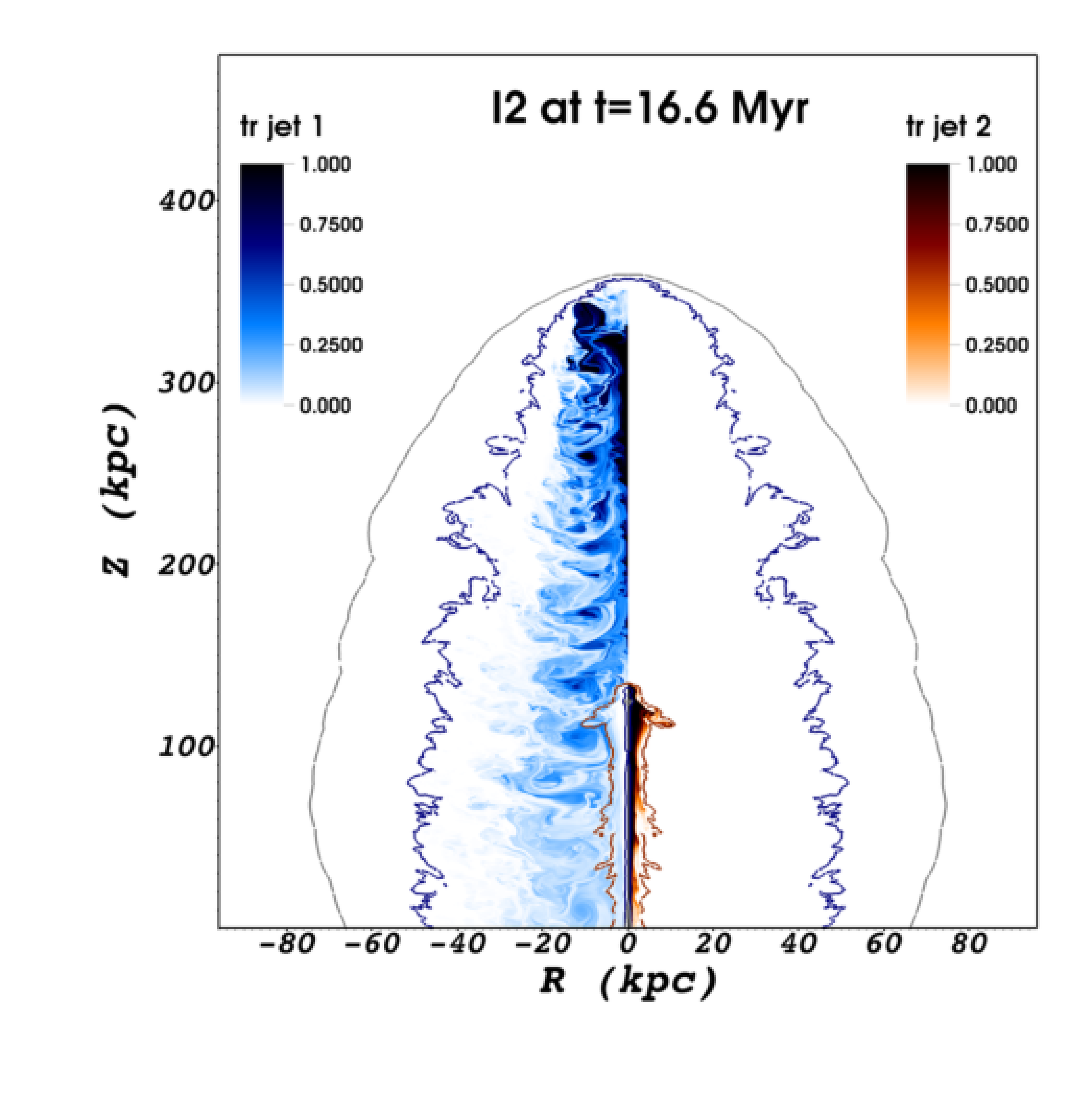}&
\includegraphics[clip=true,trim=2.5cm 2.5cm 1cm 1cm,width=0.5\textwidth]{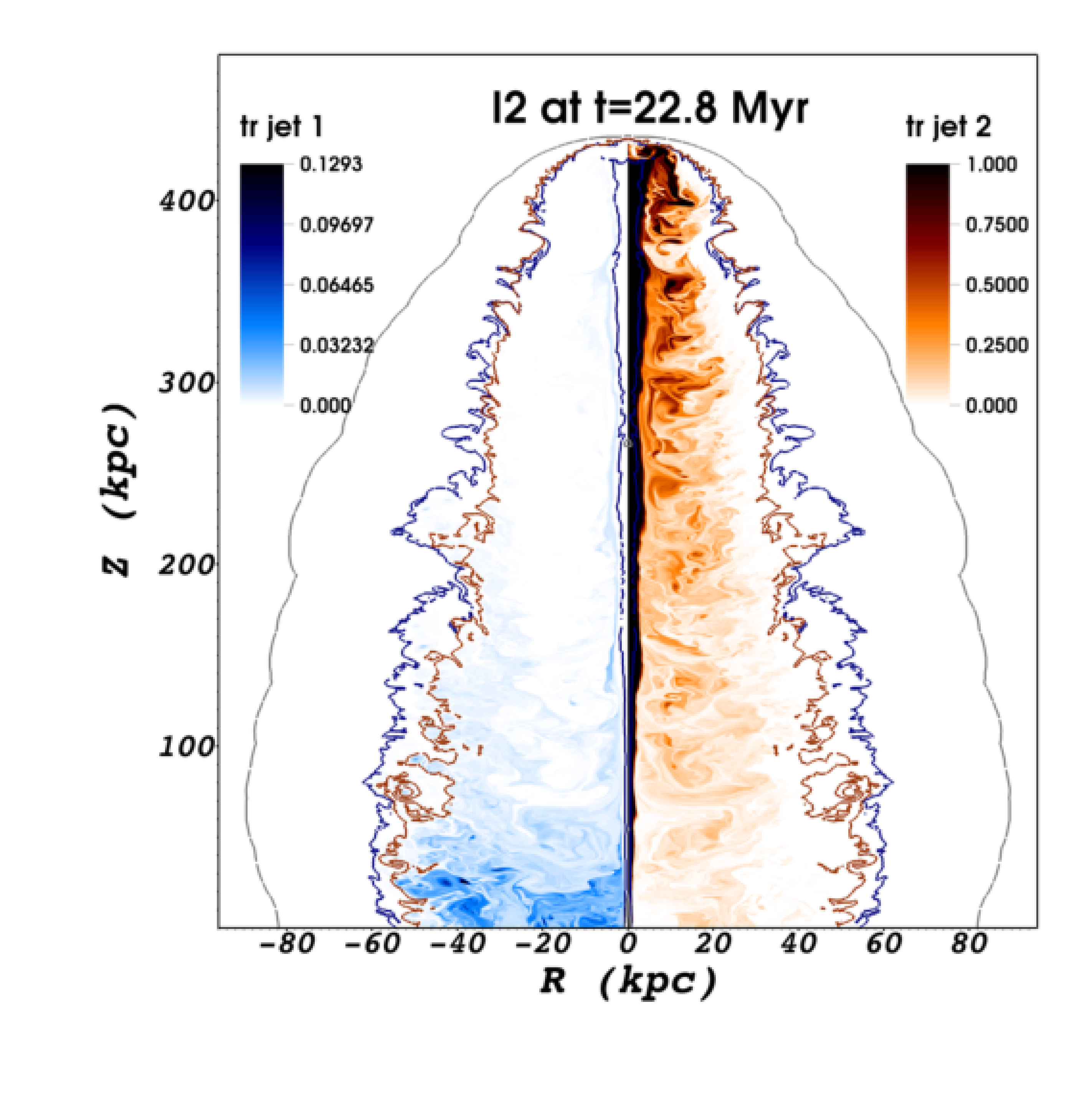}
\end{array}
$
%\end{center}
%\vspace{1.5 cm}
%\caption{Contour plots of the isothermal jet $I2$ showing the third and fourth phase of a typical episodic jet
%eruption. The $Z-$axis has been compressed by a factor of 2.5. See the caption of
%\mbox{Figure \ref{fig:A2Phase34}} for
%a full explanation.
%}
\caption{Contour plots of the isothermal jet $I2$. On the lefthand side \mbox{($t$ = 16.6 Myr)}, the plots show
the restarted jet as it is propagating completely within the $\dAM$ that was created by the initial jet.
At this time, the initial jet has not yet completely disappeared. This phase can therefore be categorized as
the overlap between phase 2 and phase 3 of a typical episodic jet eruption. On the righthand side
\mbox{($t$ = 22.8 Myr)}, the restarted jet has penetrated the forward edge of the $\dAM$ and is therefore
propagating in the $\uAM$. The plots in the top row show: in the left panel the thermal pressure
$\log(P/P_{\rm ch})$ and in the right panel the mass-weighted mixing (as described in section \ref{subsec:Mixing})
between jet spine material and jet sheath material coming from the {\em restarted jet}. The plots in the bottom
row show: in the left panel (in blue) tracer material from the initial jet
\mbox{($\theta_{\rm 1}^{\rm sp} + \theta_{\rm 1}^{\rm sh}$)} and in the right panel (in orange) material from the
restarted jet \mbox{($\theta_{\rm 2}^{\rm sp} + \theta_{\rm 2}^{\rm sh}$)}. Moreover, the gray (outer) line
contour encloses the hot cocoon that is inflated by the initial and restarted jets. In all four plots, the dark
blue (middle) line contour encloses the region that contains (shocked and unshocked) jet material from the
{\em initial} jet eruption. Finally, in all four plots the brown (inner) line contour encloses the region that
contains (shocked and unshocked) jet material from the {\em restarted} jet eruption. The $Z-$axis has been
compressed by a \mbox{factor of 2.5}.
}
  \label{fig:I2Phase34}

\vspace{10 cm}

\end{figure*}
%\end{comment}

The reduction of the impact area is the result of a strongly reduced mass discharge by the jet into the
surrounding medium through the Mach disc, which leads to a very thin layer of back-flowing material (cocoon)
around the restarted jet. The total mass discharge of jet material through the Mach disc per unit time
$\dot{M}_{\rm MD}$, as measured in the {\em Mach disc rest-frame}, equals:
\be
\dot{M}_{\rm MD} = A_{\rm jt} \: \rho_{\rm jt} \: \gamma_{\rm MD} \beta_{\rm MD} 
                 = A_{\rm jt} \: \rho_{\rm jt} \: u_{\rm MD} \; .
\label{eq:Mdot}
\ee
As before $\rho_{\rm jt}$ is proper mass density of the jet material, $A_{\rm jt}$ is the jet cross section,
$\beta_{\rm MD}$ is the velocity in units of $c$ with which the jet material enters the Mach disc,
$\gamma_{\rm MD}$ is the corresponding Lorentz factor and
\mbox{$u_{\rm MD} \equiv \gamma_{\rm MD} \beta_{\rm MD}$}. For the first jet (case 1) we find
\mbox{$u_{\rm MD} \simeq 2.81$}, while the restarted jet (case 2) has \mbox{$u_{\rm MD} \simeq 1.04$}. This leads
to:
\be
\frac{\dot{M}_{\rm MD_2}}{\dot{M}_{\rm MD_1}} = \frac{u_{\rm MD_2}}{u_{\rm MD_1}} = 0.37 \; .
\label{eq:MdotRatio}
\ee
We use the fact that both the jet cross section, as well as the proper mass density for both jets are equal and
that the energy-momentum discharge $\dot{M}$ (as measured in a given inertial frame) is approximately constant
along the entire jet axis. The total amount of mass going through the Mach disc in its rest-frame in a time
$\Delta t_{\rm MD}$ is \mbox{$\Delta M = \dot{M}_{\rm MD} \: \Delta t_{\rm MD}$}. To get the corresponding value
in the observers frame we have to take account of time dilatation:
\mbox{$\Delta t_{\rm obs} = \gamma_{\rm hd} \: \Delta t_{\rm MD}$}. Since \mbox{$\Delta M$} is a Lorentz
invariant, we find:
\be
\Delta M = \dot{M}_{\rm obs} \: \Delta t_{\rm obs} = \dot{M}_{\rm MD} \: \Delta t_{\rm MD} 
\ee
and\mbox{ $\dot{M}_{\rm obs} = \dot{M}_{\rm MD}/\gamma_{\rm hd}$}. To interpret the simulation results we
consider the amount of mass discharged in the observers frame over the time needed for each jet to travel a
length $D_{\rm co}$, which is \mbox{$t_{\rm obs} = D_{\rm co}/\beta_{\rm hd}$}. Therefore, by taking the
expression for the mass discharge \equref{eq:Mdot}, substituting the expressions for the relative velocity
\equref{eq:betaRel} and \equref{eq:gammaRel}, and accounting for the time-dilatation effects, the {\em total}
amount of jet material $\Delta M$ that passes through the Mach disc when the jet-head reaches a distance
$D_{\rm co}$ from the jet inlet is written as:
\be
\Delta M = A_{\rm jt} \rho_{\rm jt} \gamma_{\rm jt} D_{\rm co}
\left( \frac{\beta_{\rm jt}}{\beta_{\rm hd}}-1\right) \; .
\ee
The ratio of the total amount of jet material deposited through the Mach discs of the initial jet and the
restarted jet for given jet length is:
\be
\frac{\Delta M_{\rm 2}}{\Delta M_{\rm 1}} = \left(\frac{\beta_{\rm hd_1}}{\beta_{\rm hd_2}}\right)
\frac{\beta_{\rm jt} - \beta_{\rm hd_2}}{\beta_{\rm jt} - \beta_{\rm hd_1}} = 0.016 \; .
\ee
We substituted \mbox{$\beta_{\rm jt} \approx 0.95$} for the average jet bulk velocity of both jets.

In conclusion: when the jet-head of the restarted jet reaches a distance of $D_{\rm co}$ from the jet inlet, only
a fraction of \mbox{$\sim$ 1.6 per cent} is deposited through the Mach disc in comparison with the amount
deposited by the first jet over the same length. This much reduced amount of shocked jet material ultimately
leads to a very thin layer (`cocoon') of back-flowing material around the second jet.

\subsubsection{Mixing}
\label{subsubsec:Mixing}

The low discharge of back-flowing jet material from the restarted jet has an important consequence for the
behavior of mixing between spine material and sheath material in models A2 and I2, both within the jet and in the
region of back-flowing shocked jet material. Prominent vortices that are formed at the jet-head in the case of
the initial jet arise as a result of the interaction between the jet-flow and the back-flowing material that has
crossed the Mach disc. Since the back-flow associated with the second jet contains much less mass than the
back-flow from the initial jet eruption, the vortices are almost completely absent in the case of the restarted
jet. Moreover, as the restarted jet propagates approximately 16 times faster than the initial jet, it is expected
that for those vortices that do arise, the number of vortices that are shed along a cocoon of size $D_{\rm co}$
should be less by approximately the same factor.

As explained in $\SWa$, vortices play an important role in mixing the shocked spine and shocked sheath material
in the back-flow of the cocoon. Since few vortices are shed by the restarted jet, the back-flowing spine and
sheath material mix less well compared to the case of the initial jet. This is also seen in the top left panel of
\mbox{Figure \ref{fig:I2Phase34}}. In the case of the restarted jet, the jet-cocoon coupling is very weak. The
internal shocks are less strong and this results in a significantly more stable transverse structural integrity
for the isothermal jet, as well as the isochoric jet. This can for example be seen in the bottom left panel of
\mbox{Figure \ref{fig:ZcutsA2}}, where the centre of the restarted jet remains dominated by jet spine material
almost up to the Mach disc. For the isochoric jet $A2$, the increase in transverse structural integrity of the
second jet compared to the first jet is also clearly visible in \mbox{Figure \ref{fig:RcutsA2}}.

In \mbox{Figure \ref{fig:TimePlots}}, phase 3 is enclosed by the dashed line in the middle (at
\mbox{$t = 16.0$ Myr}) and the dashed line on the right at (\mbox{$t \sim 18$ Myr}). In these plots, it is
immediately seen that no prominent hotspots emerge, as long as the second jet propagates inside the $\dAM$:
the threshold of the pressure is exceeded (almost) nowhere along the jet-axis and the effective polytropic index
and the temperature do not become relativistic. Only just before the jets break out of the $\dAM$ do they
start to decelerate, and a hotspot re-appears.

%\begin{comment}
%%% Contour plots of the $I2$ and $A2$ jets, showing mixing between sp1 and sh1 (right) and jt1 and jt2 (left).
\begin{figure*}
%\begin{center}
$
\begin{array}{c c}
\includegraphics[clip=true,trim=2.5cm 2.5cm 1cm 1cm,width=0.5\textwidth]{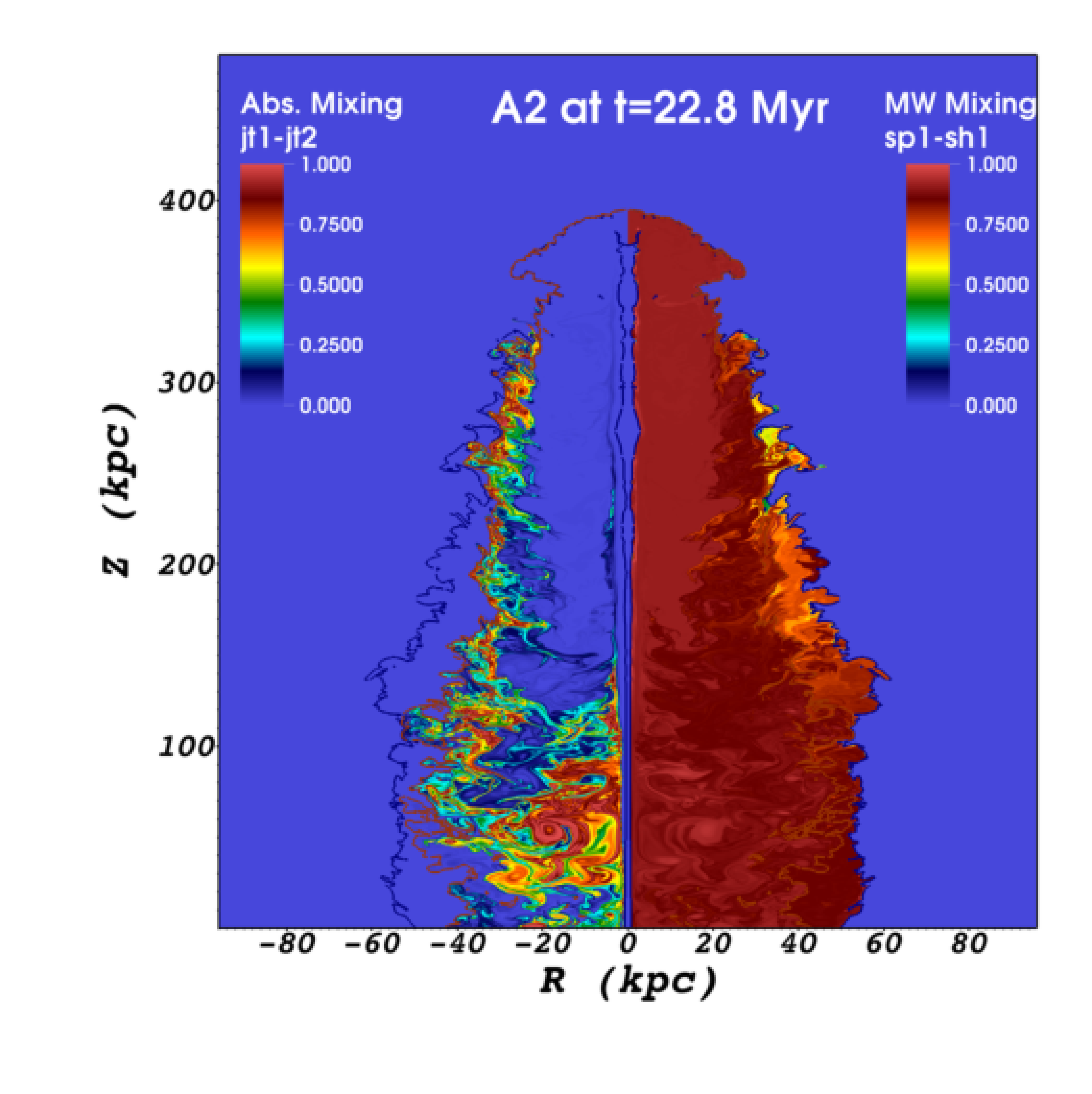}&
\includegraphics[clip=true,trim=2.5cm 2.5cm 1cm 1cm,width=0.5\textwidth]{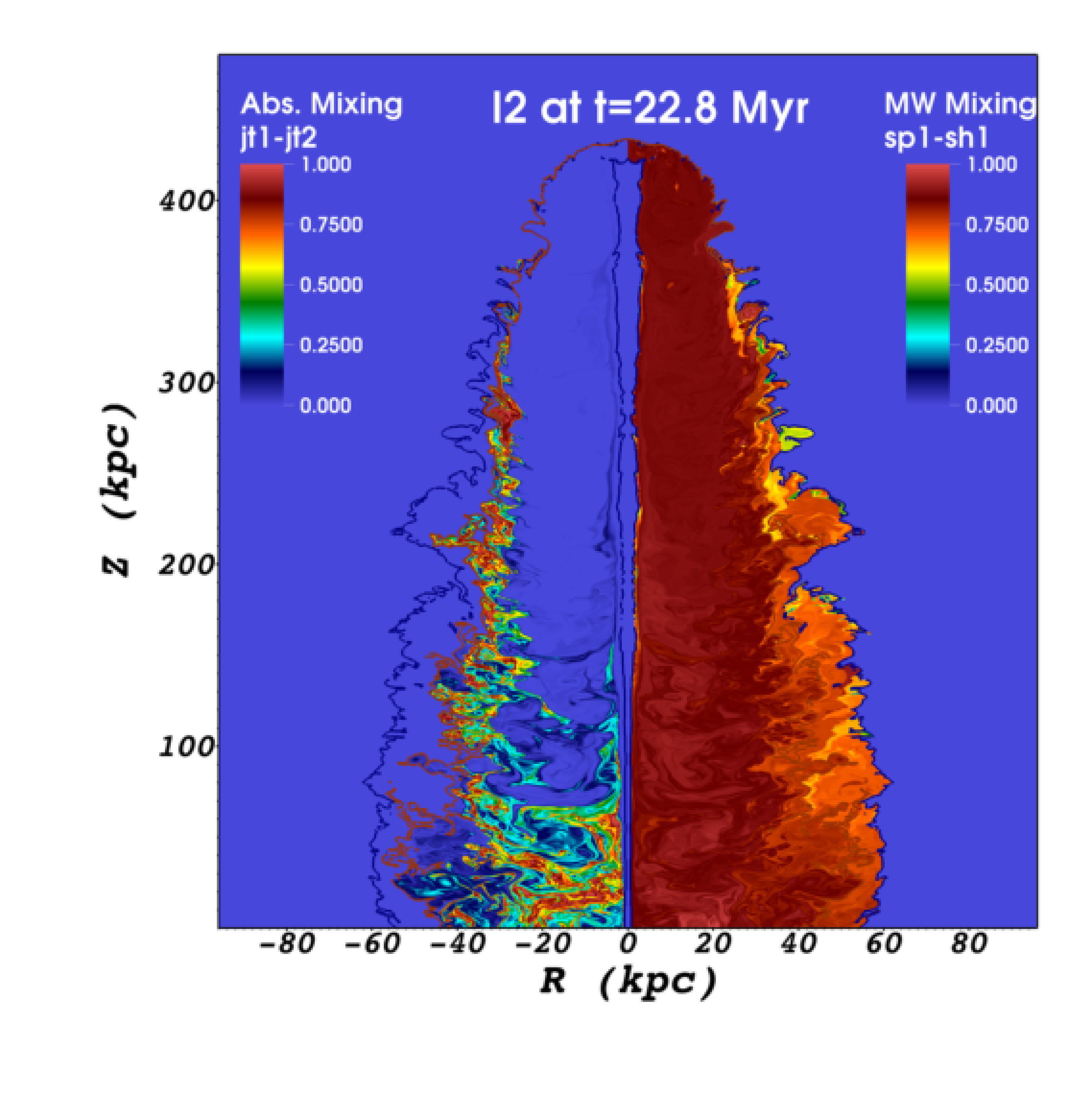}\\
\end{array}
$
%\end{center}
%\vspace{1.5 cm}
\caption{Contour plots of the isochoric jet $A2$ (left) and the isothermal jet $I2$ (right) at the end time of
simulation, \mbox{$t_{\rm tot} = 22.8$ Myr}. The left panels of the contour plots show the absolute mixing
between material from the initial jet \mbox{($\theta_{\rm 1}^{\rm sp}+\theta_{\rm 1}^{\rm sh}$)} and the
restarted jet \mbox{($\theta_{\rm 2}^{\rm sp}+\theta_{\rm 2}^{\rm sh}$)}. The right panels show the mass-weighted
mixing between spine material $\theta_1^{\rm sp}$ and sheath material $\theta_1^{\rm sh}$ from the {\em initial}
jet eruption as it has evolved at the final simulation time. The dark-blue and brown contours have the same
meaning as in \mbox{Figure \ref{fig:I2Phase34}}. These contour plots show two notable features:
1. most of the material from the initial jet eruption is pushed downwards or (radially) outwards by renewed
shocked jet material from the restarted jet. This is where most mixing between material from the initial jet and
material from the restarted jet takes place. 2. Near the jet-head, where the back-flow of the restarted jet
is strongest, the small amount of material from the initial jet that is still present has mixed into a near
perfectly homogeneous mixture. The $Z-$axis has been compressed by a factor of 2.5.
}
  \label{fig:A2I2Phase4}

\end{figure*}
%\end{comment}

\subsection{Phase 4: jet break-out and further propagation}

In the fourth and final stage, the restarted jet has broken out of the $\dAM$ and is now propagating in the
$\uAM$.  Just before the jet-head of the restarted jet reaches the front end of the $\dAM$, it briefly encounters
the region that contains purely shocked IGM material from the initial jet eruption. The mass density in that
region is higher than that of the $\uAM$. Once through this region of compressed gas, the jet-head advances
through a medium with the same properties as the $\uAM$ encountered by the first jet. Therefore, after transients
have died down, the jet behaves very similar to the first jet, see for instance
\mbox{Table \ref{tab:EruptionPhase}}.

\subsubsection{Jet-head advance speed and cocoon formation}

As soon as the jet-head of the restarted jet runs into the denser shocked IGM material it decelerates. From
that moment on, the termination shock and the forward bow shock increase in strength. As a result, the jet
material that passes through the Mach disc is now shocked to relativistic temperatures. This causes the hotspots
to re-emerge. 

The strong back-flow of shocked jet material is re-established as soon as the restarted jet runs into the denser
medium and slows down. This material quickly fills a large fraction of the old cocoon with shocked jet material
from the second jet. The mixing between the shocked back-flowing spine and sheath material increases. Moreover,
the increase in back-flow causes strong pressure fluctuations along the jet axis that lead to internal shocks
within the jet, as described in more detail in $\SWa$. The transverse structural integrity of the isochoric jet
$A2$ is affected by these internal shocks.

The breakout from the $\dAM$ and further propagation in the $\uAM$ of the second jet is also shown in
\mbox{Figure \ref{fig:TimePlots}} between the dashed line on the right (at \mbox{$t = 18$ Myr}) and the right
boundary (at \mbox{$t = 22.8$ Myr}). The restarted jet in phase 4 evolves in a very similar fashion as the
initial jet in phase 1, as can be seen in the panels for pressure and effective polytropic index, as well as from
the jet-head propagation speed (see \mbox{Table \ref{tab:EruptionPhase}}). The onset of the strong back-flow of
shocked jet material that occurs when the second jet breaks out of the $\dAM$ can be recognized in the pressure
panels of \mbox{Figure \ref{fig:TimePlots}}. There, a declining straight feature stretches from the point of
breakout to the bottom right. This is the internal shock that runs downwards through the cocoon as a result of
the onset of the strong back-flow.

\subsubsection{Mixing}

The dark-blue contour in \mbox{Figure \ref{fig:I2Phase34}} marks the region within the cocoon that contains
material from the initial jet. To some extent it has mixed with the shocked IGM. Directly outside of this
contour, but still inside the initial cocoon, lies the region of purely shocked material from the IGM. The brown
contour in these plots has a similar meaning, although now for the restarted jet. It marks the boundary that
separates the region containing material from the restarted jet (mixed to some extent with the $\dAM$) from the
region that does not contain any material from the restarted jet. These plots show that the brown contour
largely overlaps with the dark-blue contour, especially near the jet-head where the material of the restarted
jet has had the chance to `catch up' with the material from the initial jet. The fact that these contours tend
to overlap suggests that the evolution of the shocked jet material from the restarted jet is strongly influenced
by the structures that were created by the initial jet. This opens up the possibility that old radio
lobes/structures that were created by the initial jet can be re-energized through either Fermi-II acceleration of
old electrons, or the injection of new relativistic electrons.

\mbox{Figure \ref{fig:A2I2Phase4}} shows two additional forms of mixing at the final time of simulation
\mbox{$t=$ 22.8 Myr}. In the left panels of the contour plots, it shows the absolute mixing between material from
the initial jet \mbox{($\theta_1^{\rm sp}+\theta_1^{\rm sh}$)} and that of the restarted jet
\mbox{($\theta_2^{\rm sp}+\theta_2^{\rm sh}$)}. In the right panels, it shows the mass-weighted mixing between
spine material $\theta_1^{\rm sp}$ and sheath material $\theta_1^{\rm sh}$ of the {\em initial} jet. Two
notable features can be seen in these contour plots. The first is that near the jet-head, where the back-flow of
new shocked jet material is strongest, there is very little mixing between material from the initial jet and the
restarted jet. This means that the new back-flowing/high pressure jet material pushes the older material away
from the jet-head, either radially outwards or downwards. 

The second notable feature is that the strong back-flow of the restarted jet does not entrain all of the material
from the initial jet. Rather, a small fraction \mbox{$< 0.1 \%$} is left behind and all inhomogeneities between
$\theta_{1}^{\rm sp}$ and $\theta_{1}^{\rm sh}$ are washed out, so what is left is a nearly perfect homogeneous
mixture of spine and sheath material from the first jet. Note, however, that at the same time, this region shows
the {\em least} homogeneity between back-flowing spine and sheath material from the restarted jet (see upper
right panel of \mbox{Figure \ref{fig:I2Phase34}}).

\subsection{Free-free emission from the cocoon}

%\begin{comment}
%%% Free-free emission!
\begin{figure*}
%\begin{center}
$
\begin{array}{c c c}
\phantom{} &
\includegraphics[clip=false,trim=0cm 0cm 0cm 0cm,width=0.4\textwidth]{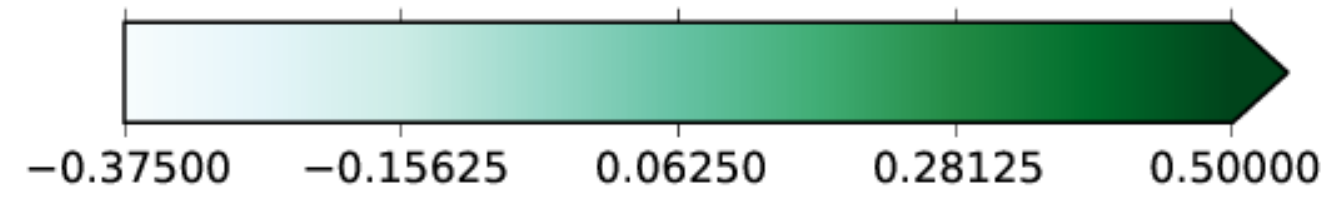}&
\phantom{} \\
\includegraphics[clip=false,trim=0cm 0cm 0cm 0cm,width=0.25\textwidth]{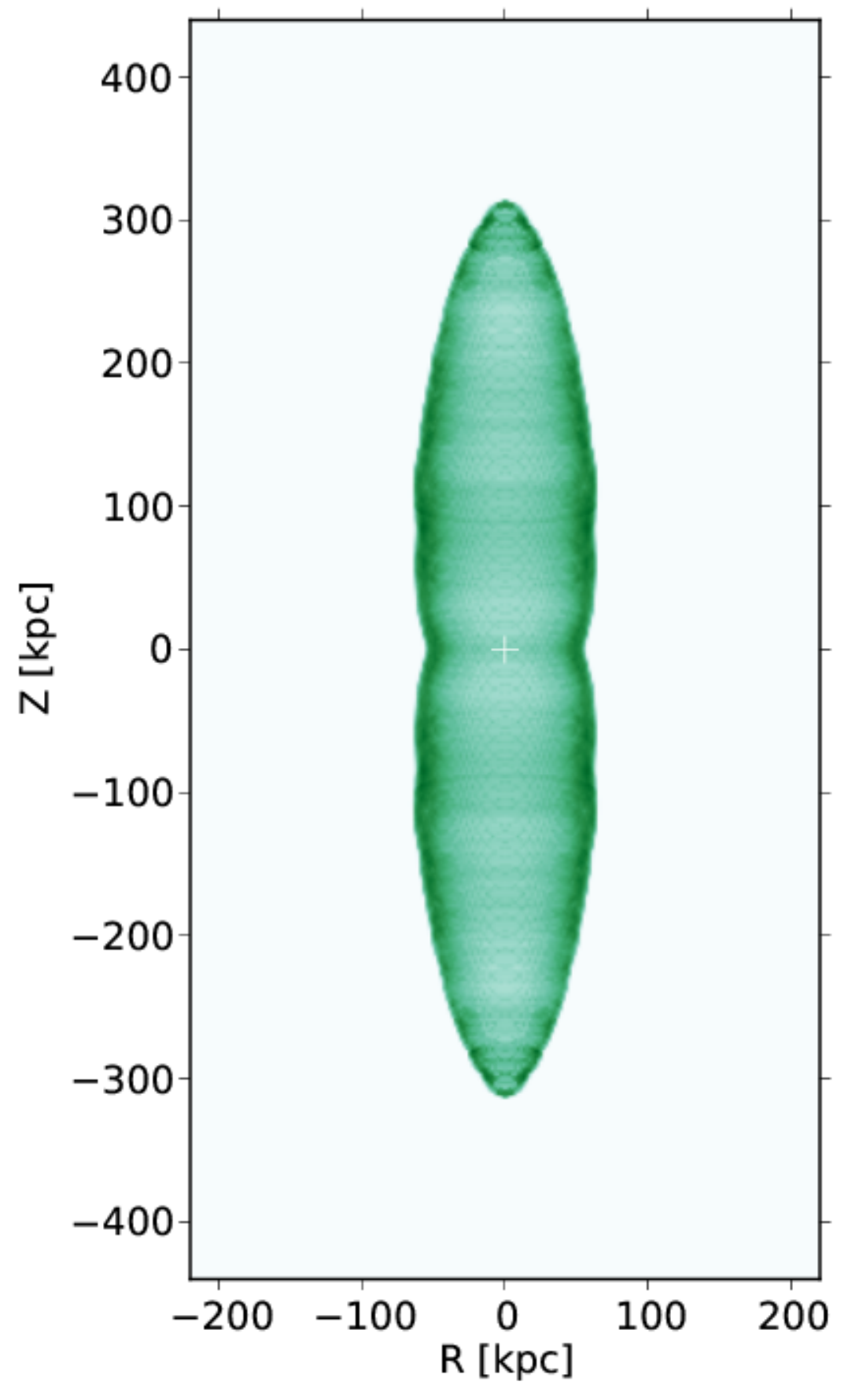}&
\includegraphics[clip=false,trim=0cm 0cm 0cm 0cm,width=0.25\textwidth]{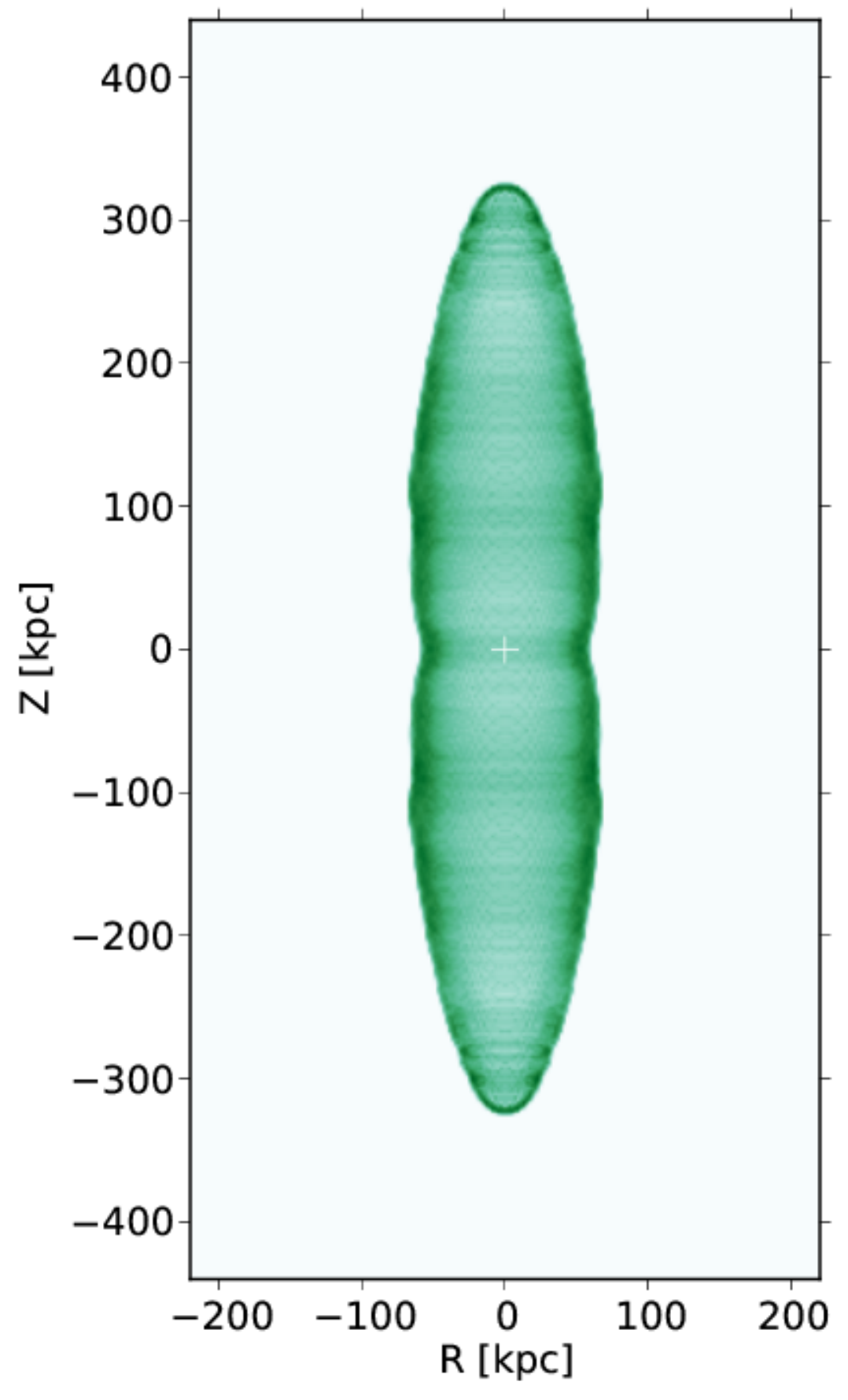}&
\includegraphics[clip=false,trim=0cm 0cm 0cm 0cm,width=0.25\textwidth]{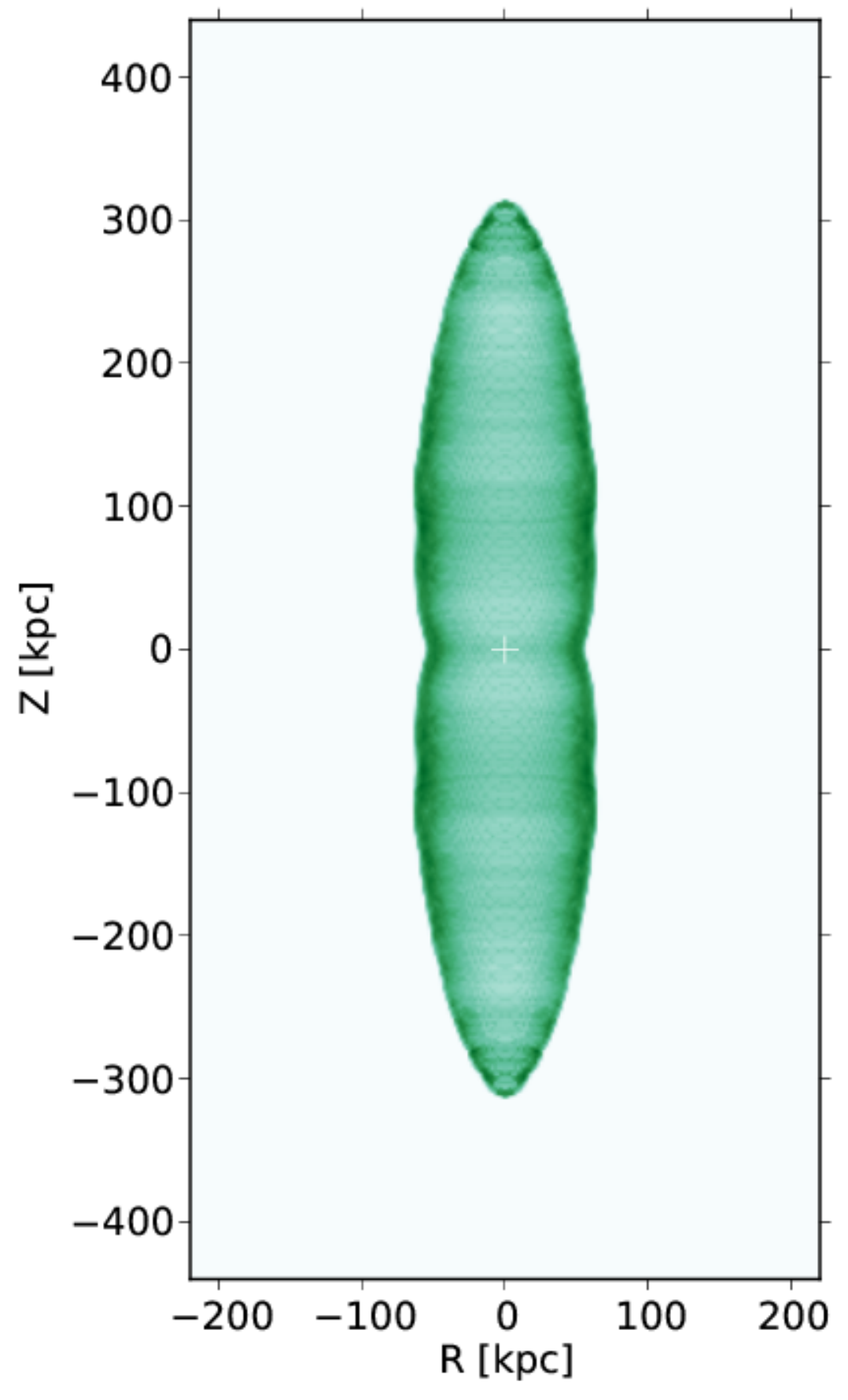}\\
\phantom{} &
\includegraphics[clip=false,trim=0cm 0cm 0cm 0cm,width=0.4\textwidth]{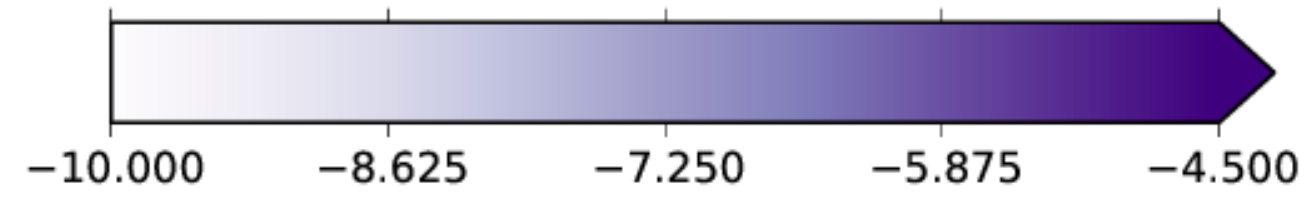}&
\phantom{} \\
\includegraphics[clip=false,trim=0cm 0cm 0cm 0cm,width=0.25\textwidth]{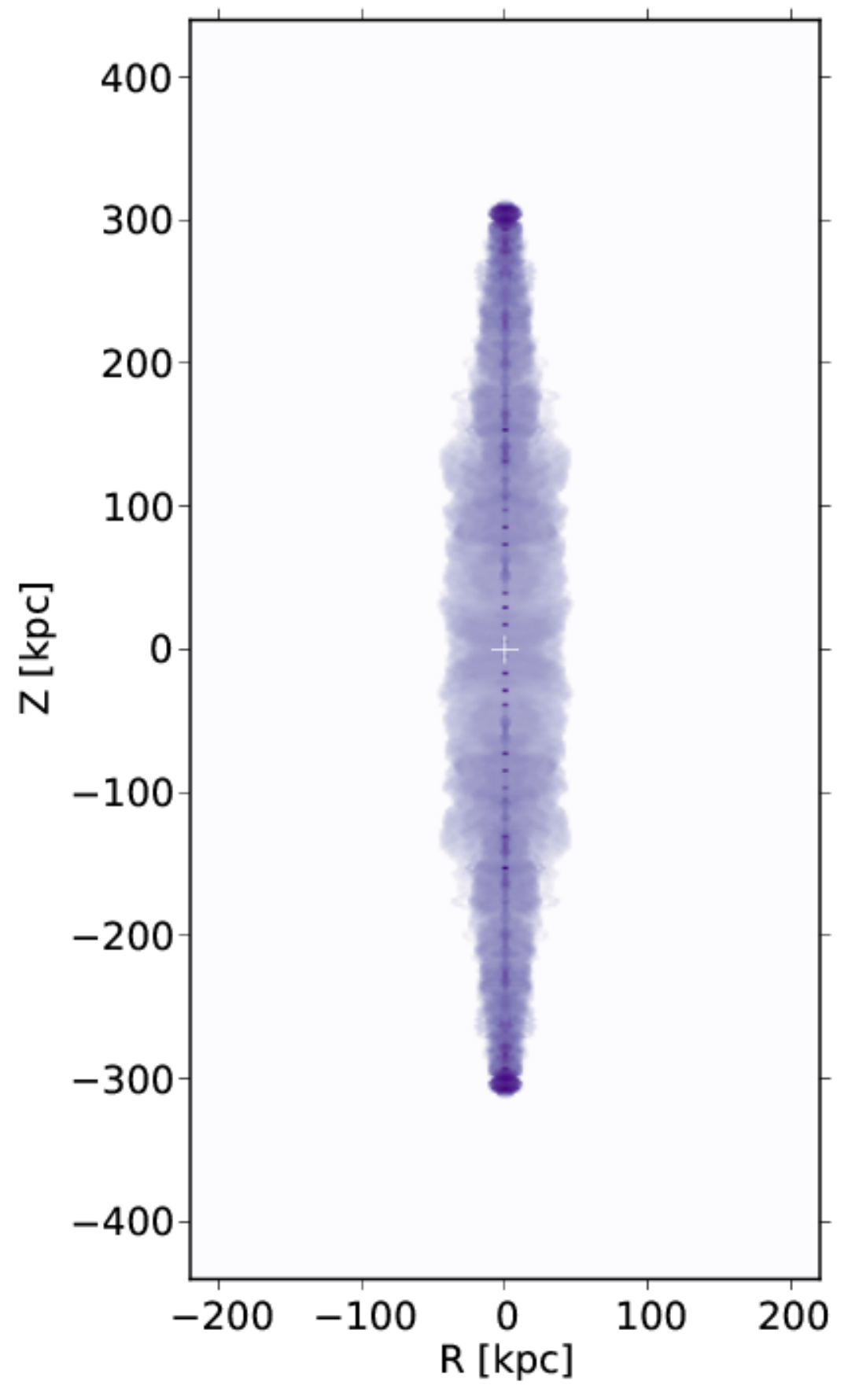}&
\includegraphics[clip=false,trim=0cm 0cm 0cm 0cm,width=0.25\textwidth]{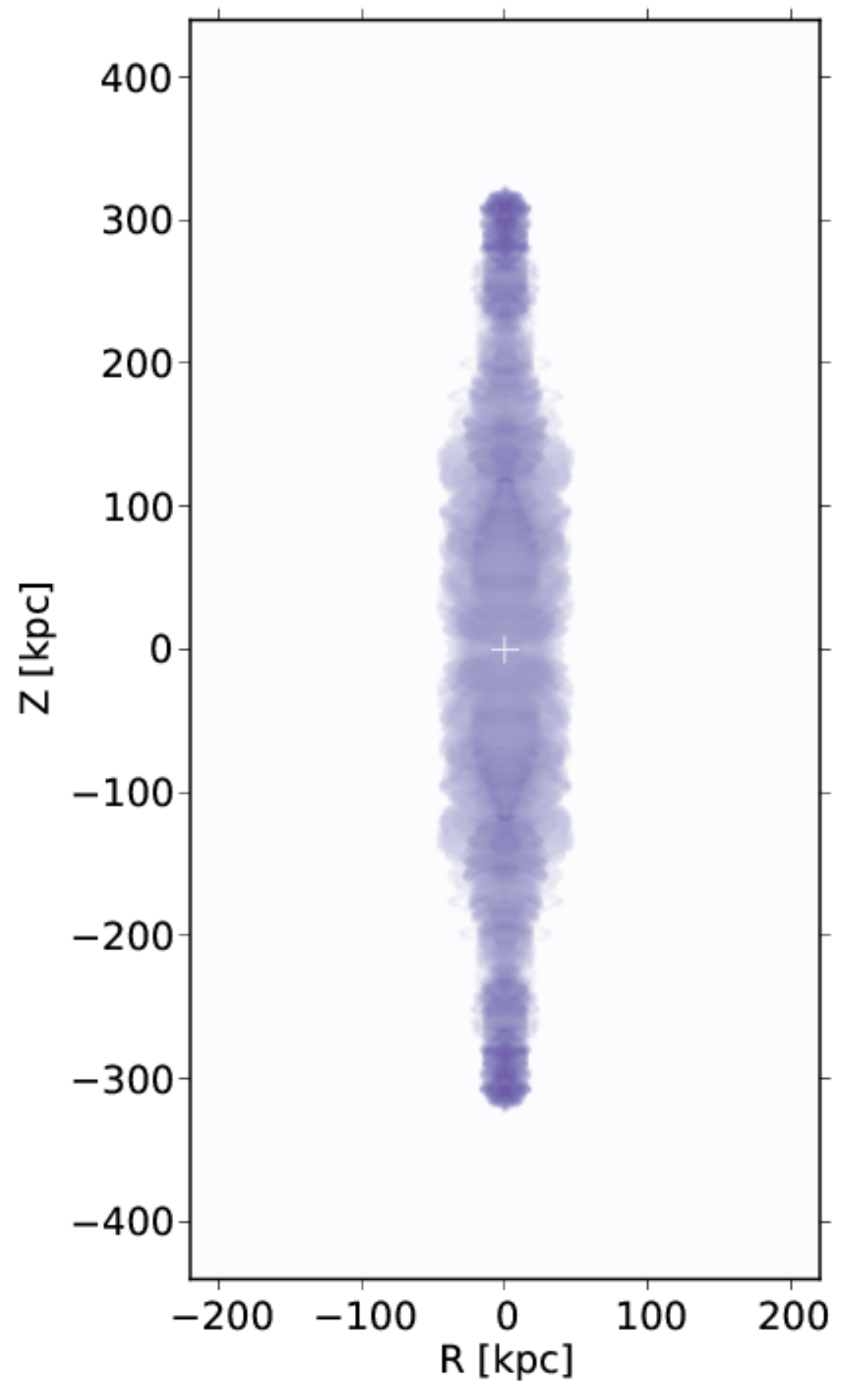}&
\includegraphics[clip=false,trim=0cm 0cm 0cm 0cm,width=0.25\textwidth]{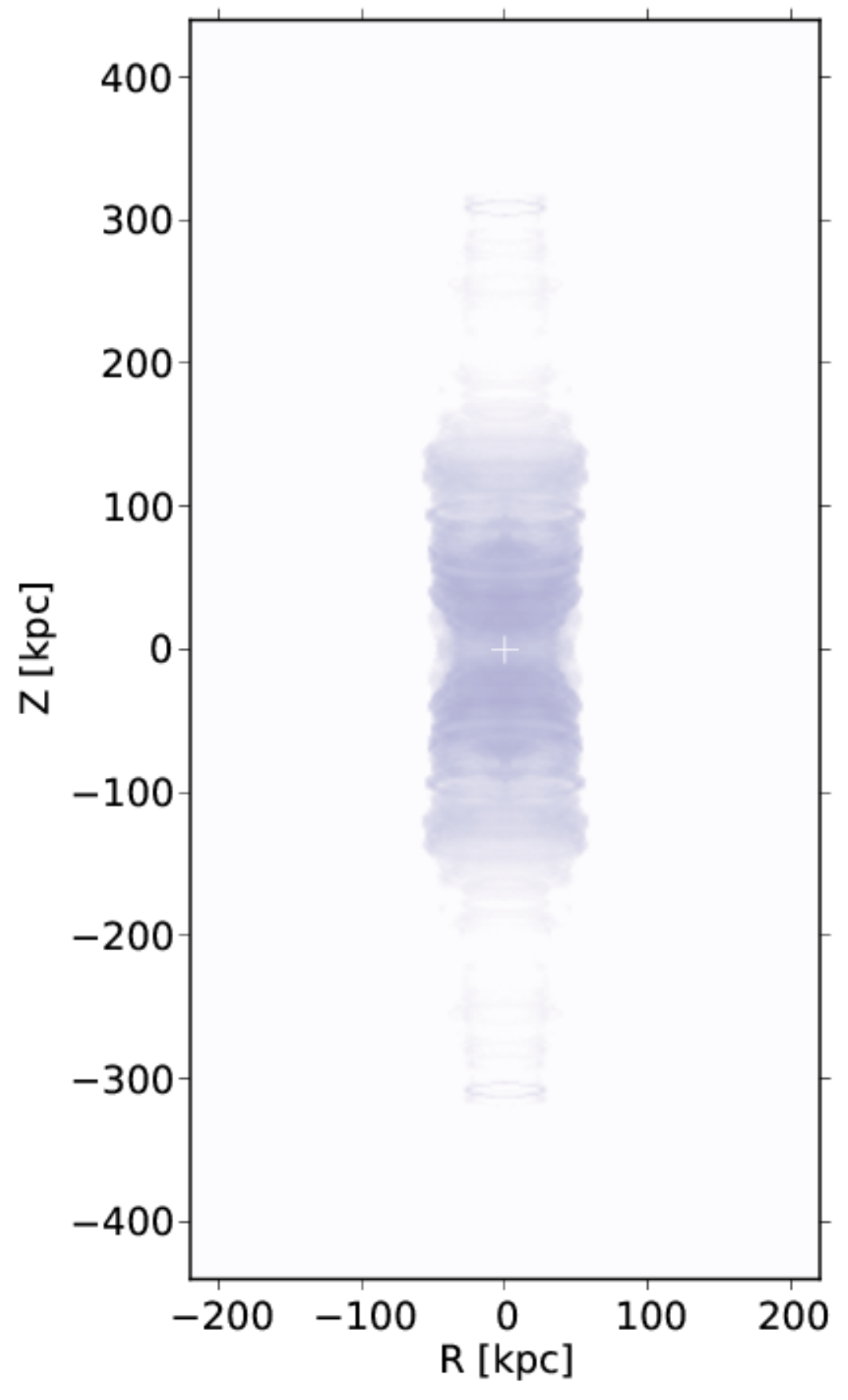}
\end{array}
$
%\end{center}
%\vspace{1.5 cm}
\caption{Free-free emission for the isochoric jet ($A2$) in phase 1 at \mbox{$t = 15.3$ Myr} (left panels), in
phase 3 at \mbox{$t = 16.6$ Myr} (middle panels) and in phase 4 at \mbox{$t = 22.8$ Myr} (right panels). The
plots are constructed by rotating the $2D$ plots along the jet axis on to a $3D$ grid, after which the emission
is integrated along the line of sight. The jets are reflected in the plane \mbox{$Z = 0$}. The viewing angle
is $80^{\circ}$ to the line of sight. The emission is shown on a logarithmic scale in arbitrary units. The top
panels (in green) show the total free-free emission coming from jet material, as well as the ambient medium. The
lower panels (in purple) show the contribution to the free-free emission from jet material alone. The units of
the total free-free emission and the contribution from the jet material are equal.
}
  \label{fig:FreeFree}

\end{figure*}
%\end{comment}

The fact that the material left by the first jet (and later by the second jet in phase 4) in the broad cocoon is
very hot and very tenuous means that this material can be a source of free-free emission, with a cooling time
similar to or even larger than the age of the source. This means that in X-rays the cocoon is a long-lived
feature that remains visible long after the jet (or jets) causing it have been turned off.

\mbox{Figure \ref{fig:FreeFree}} shows the free-free emission (surface brightness) in the optically thin limit
from the jet and cocoon in phases 1 (left panel, \mbox{$t = 15.3$ Myr}, just before the first jet is shut off),
phase 3 (middle panel, \mbox{$t = 16.6$ Myr}) and phase 4 (right panel, \mbox{$t = 22.8$ Myr}). The top row shows
the total free-free emission, calculated using an emissivity equal to:
\be
\epsilon_{\rm ff} \propto n^2 \: \sqrt{P / n} \; .
\ee
It therefore includes the contribution from the jet, as well as the contribution from the shocked intergalactic
medium. The bottom row, on the other hand, just shows the contribution of material from the first jet to the
free-free emission. It is calculated using a weighted emissivity:
\be
\epsilon_{\rm ff-jet_1} \propto
                        \left[(\theta_{1}^{\rm sp} + \theta_{1}^{\rm sh} ) \: n\right]^2 \: \sqrt{P / n} \; .
\ee
From these plots, it can clearly be seen that the regions of lowest brightness correspond to the regions that
contain most of the jet material. In other words, the low-density regions that are inflated by the shocked jet
material result in cavities in the X-ray emission. We also note that in the lower left panel of
\mbox{Figure \ref{fig:FreeFree}}, the internal shocks along the jet axis, as well as the hotspots can clearly
be seen. However, since the contribution of the jet material is still relatively small compared to the total
emission, these shocks do not show up in the plots of the total emission.

The propagation of the second jet inside the $\dAM$ is also shown in the centre panels. A faint signature of the
restarted jet can be seen in the lower centre panel. The material of the first jet is compressed by the (weak)
bow shock of the second jet. However, this compression is so weak that it does not show up in the total
contribution of free-free X-ray emission.

Finally, the lower right panel shows that the X-ray emission coming from material of the first jet is displaced
toward the centre of the image. So for a radio galaxy in the fourth phase of double-double evolution, most of the
material from the earlier eruption remains closer to the parent galaxy and away from the new jet-head and hotspot.

\section{Discussion and link with observations}
\label{sec:Discussion} 

  \subsection{Linking various phases to observations of radio galaxies}
\label{subsec:Linking} 

As the simulations in this paper reveal, episodic jets result in a number of distinct features that characterize
the various phases of source evolution. A number of these features appear to correspond with what is observed
in a number of different radio galaxies. We briefly list them below:

\subsubsection{Phase 1}

Continuously driven relativistic- and under-dense jets have been studied in detail $\SWa$, and again here for
phase 1. The main results are for our choice of parameters: [1] a low jet-head advance speed [2] a strong
back-flow of shocked jet material [3] strong jet-cocoon coupling resulting in strong internal shocks along the
jet axis [4] a relatively thick cocoon [5] formation of hotspots and [6] a large effective impact area. It should
be noted that these features are to some extent the result of our initial parameter choice. For example, jets
that are strongly over-dense compared to the ambient medium propagate almost ballistically, and evolve in a
significantly different fashion.

\lskip
A list of \mbox{$\sim 30$} powerful FRII radio galaxies that seem to have been driven at a roughly constant
mechanical luminosity over the entire lifetime of the source can be found in \citet{ODea2009}. Therefore, these
sources are representative for radio galaxies in phase 1. In that study, the authors estimate a number of parameters
for the jet, the hotspots and the radio lobes. These include the source age, distance from the central engine to
the hotspots, jet-head propagation speed, total jet power, pressure in the radio lobes, and density of the
ambient medium. For many of these sources these parameters are in close agreement with the values used in or
obtained from the simulations of this study.

\subsubsection{Phase 2}

In the second phase the accretion flow that is feeding the central engine of a radio galaxy is temporarily halted
or changes its character, so that the jets are switched off.

As the simulations in this paper show, the time it takes for an AGN jet with a typical size of
\mbox{$\sim 10^2 - 10^3$ kpc} to completely disappear is approximately \mbox{1.2 Myr}, a very short time compared
to the typical age of such a radio galaxy (which is of order \mbox{$\sim 10 - 100$ Myr}).

Therefore, observations of extragalactic radio sources in this phase statistically favor the situation where the
trailing end of the initial jets have crossed the Mach disc, so that the bulk jet flow and the associated strong
shocks (Mach disc and bow shock) have completely disappeared.

What remains after the jets have disappeared is a long-lived, over-pressured and under-dense cocoon. This cocoon
can be identified with the observed X-ray cavities around some radio galaxies in large galaxy clusters. Examples
of such cavities are the X-ray bubble around Cygnus A (e.g. \citealt{Carilli1996}), or the X-ray super-cavities
in the Hydra A cluster, which seem to be a result of a series of jet eruptions over the past
\mbox{$200 - 500$ Myr} \citep{Wise2007}.

It is challenging to catch an AGN in the act of switching off its jets. The reason is that the time it takes for
the jets to completely disappear after switch-off is typically short compared to the total age of the source.
However, \citet{Tadhunter2012} report a rare example of a radio loud/radio quiet double AGN system,
\mbox{PKS 0347+05}. Observations in optical, infrared and radio frequencies suggest that both AGNs have been
triggered by a major galaxy merger that took place within the last \mbox{100 Myr}. The powerful FRII source with
extended radio lobes and hotspots shows only weak, low ionization emission line activity near the nucleus. This
behavior can be explained by a rapid decline in nuclear AGN activity within the last \mbox{$10^6$ yr}. It
suggests that the central engine of the AGN has recently switched off its jets, while decrease in AGN activity
and jet interruption has yet to affect the radio lobes and hotspots of this radio galaxy.

\subsubsection{Phase 3}

In phase 3 a restarted jet propagates at high speed through the tenuous cocoon left by the earlier jet.
A typical DDRG, such as \mbox{B1545-321}, is in phase 3 according to criteria suggested by the results of
this paper. For example (where we use the results of \citealt{Safouris2008} in the remainder of this Section):
\begin{itemize}
\item The age of B1545-321 is estimated to be \mbox{$(0.3 - 2) \times 10^8$ yr}, while the total length of the
source is approximately \mbox{1 Mpc}. The interruption time for B1545-321 is estimated to be just a few percent
of the duration of the initial jet eruption.

\item
The outer radio lobes of B1545-321 are no longer fed by material from the first jet. It is believed that their
outward motion, away from the parent galaxy, stopped \mbox{$1.8 \times 10^5$ yr} ago. The distance of the inner
hotspots to the AGN suggests that the second pair of jets started at a time when the first pair of jets had not
completely disappeared.

\item 
No evidence for bow shocks associated with the inner hotspots of the restarted jet has been found in B1545-321. 
Our results show similar behavior: first of all, the jet-heads of the restarting jets are very small in size
compared to those of the initial jets. As a result, the spatial separation between the Mach disc and the bow
shock is also small so that a distinction between hotspot and bow shock will be harder to make observationally. 
Secondly, the Mach numbers of both Mach disc and bow shock of the restarted jet are much smaller than for the
initial jet, by a factor 1.45 and 15.0 respectively for the jets in this research. Since the strength of a shock
is generally determined through its proper Mach number squared ($\mathcal{M}^2$), they are significantly weaker
shocks (in particular in the case of the bow shock) that may be more difficult to detect in radio observations.

\item
As for the Mach number of the Mach disc of the restarted jet, \citet{Safouris2008} give a dynamical estimate that
depends on the cocoon pressure, the pressure of the hotspots, and viewing angle. By letting the hotspot pressure
vary between 1 and 10 times an estimated pressure minimum (synchrotron equipartition model), they find a Mach
number between \mbox{$5 \lesssim \mathcal{M}_{\rm MD} \lesssim 15$}. This is in excellent agreement with the
results from our simulations.

\item
In B1545-321, the jet-head advance speed depends on the viewing angle, believed to be in the range
\mbox{$70^{\circ} \lesssim \psi \lesssim 80^{\circ}$}. In that case, the jet-head advance speed of the restarted
jet, inferred from the spectral ages of different radio emitting regions and their distances to the nucleus,
varies between \mbox{$0.3 \lesssim \beta_{\rm hd_2} \lesssim 0.6$}. These values are similar (to within a factor
$\sim 2$) to those found in our simulations.
%The different values can be easily attributed to small differences in the fundamental parameters (e.g.
%luminosity, density ratio etc.) of the jets.
\end{itemize}

\subsubsection{Phase 4}
 
The restarted jet enters phase 4 when it has completely traversed the cocoon left by the first jet. At that point
a strong back-flow is once again triggered, leading to the formation of an extensive cocoon around the jet-head. 
When the restarted jet is active for the time it takes for the back-flowing material to reach the jet base, most
of the remnant cocoon will have been filled with renewed shocked jet material. From that point onwards, it will
be difficult to detect any jet material originating from the first jet eruption: most of that material will have
been pushed outwards, away from the jet-head, where it eventually mixes strongly with the other constituents. In
that case, the only signs of an earlier jet event having taken place might be in the form of the older bow shocks
propagating in the $\uAM$, possibly showing up as the earlier mentioned X-ray cavities.

In the work of \citet[][]{Chon2012}, the authors report a cavity in the X-ray emission that coincides with an
excess in radio emission in the cocoon of Cygnus A near the plane of the parent galaxy. They find that the
spectral age and buoyancy time of the cavity lies in between 1 and 2 times the age of the current Cygnus A jets.
They suggest therefore a scenario where the cavity was created in an earlier jet eruption that took place
more than \mbox{$\sim$ 30 Myr} ago.

The suggestion that the typical FRII radio galaxy Cygnus A recently went through an episodic event fits well the
context of our simulations. Cygnus A shows two very collimated jets with few internal shocks, strong hotspots and
extended radio lobes. From our simulations, we find that collimated jets with a strong radial structural
integrity, prominent hotspots and extended lobes of back-flowing jet material all occur when the restarted jets
are propagating in the earlier stages of phase 4, when the strong back-flow driven by the second jet has not yet
reached the lower regions near the jet inlet.
%\end{description}

  \subsection{Possible extensions of this study}
\label{subsec:Extensions}

\subsubsection{Fanaroff-Riley Class: FRI vs. FRII}

We investigated the case of jets with a luminosity typical for FRII radio galaxies and chose the conditions
of the ambient medium equal to those inferred for cluster environments. We find that jets propagating in such an
environment strongly disturb the $\uAM$ so that a restarting jet (initially) propagates in a completely different
environment. The less powerful FRI jets, on the other hand, are thought be decelerated at much smaller distances
from the central engine. Instead of prominent hotspots, more diffusive radio plumes are formed. Recently,
\citet*{Perucho2013} has simulated jets with typical properties of FRI jets, clearly showing the deceleration at
small scales and the lack of a prominent hotspot.  In that case, episodic jet activity might lead to
very different results for jet propagation, stability, mixing effects and morphology than the results found in
this paper. Such a study would probe a very different class of radio galaxies and would therefore contribute to
our understanding of episodic jet behavior and radio galaxy evolution in general.

\subsubsection{Numerical approach: boundary conditions}

The jet simulations in this paper make use of open outflow boundary conditions. This choice is particularly
useful for studying the propagation and large-scale evolution of one side of a radio galaxy, i.e. an individual
jet. This has been the main focus in $\SWa$ and this paper. In a different scenario one could better track the
full evolution of back-flowing jet material from both the initial and the restarting jet in a full 2-sided radio
galaxy by choosing reflective boundary conditions at the lower boundary. These will have an effect on the cocoon
near the plane where the jets are injected (mainly a thicker cocoon). There, it is expected that the older cocoon
material near the parent galaxy will be displaced radially outwards. In the plot of the free-free emission in
\mbox{Figure \ref{fig:FreeFree}}, we showed that regions containing (shocked) jet material contribute only very
little to the total free-free emission and show up as cavities in the synthesized X-ray plots. Therefore, it is
expected that choosing reflective boundary conditions will result in larger X-ray cavities in the free-free
emission plots near the parent galaxy, similar to the results of \citet{Chon2012}.

\subsubsection{Limitations of axisymmetry}

Finally, we mention the use of axisymmetry in this paper. It has the advantage of capturing many important
features of jet evolution, while the computational resources can be managed fairly well. The disadvantage, on the
other hand, is that no instabilities associated with the third direction, leading to more realistic asymmetries
in the jets and radio lobes are able to develop. Simulations of jets in full $3D$ often lead to strong
asymmetries in the cocoon and wiggling of the jet at larger distances. This effect gets even stronger with an
inhomogeneous or clumpy ambient medium (see for example \citealt{Mendygral2012} or \citealt{Porth2013}). We
speculate that performing the simulations in this paper in full $3D$ will lead to an increase in mixing between
the various constituents, while the radial structural integrity of the jets will decrease as the jets evolve in
the various stages of the episodic jet outburst. In particular, the $\dAM$ left by the first jet might have
developed strong asymmetries before the second jet is injected.

    \subsubsection{Continuation of this work}
\label{subsubsec:Continuation}

In a follow-up paper, we will model synchrotron emission, based on the same hydrodynamic simulations of the
jets in this paper. There, the aim will be to create images that have a close resemblance with a DDRG such as the
test case B1545-321, in terms of intensity contrasts, Doppler boosting and dimming, appearance of the hotspots,
etc. We will also study how viewing angle affects the appearance of the source. Moreover, we will consider and
compare a number of different emission mechanisms. We will show synchrotron maps during the four different
phases of episodic activity. And finally, by making use of the separate tracers for each jet constituent, we
will be able to show the separate contributions from the different jet constituents to the synchrotron surface
brightness.

%\clearpage
%\newpage

\section{Conclusions} \label{sec:Conclusions}

In this paper, we simulated episodic jet activity for relativistic and under-dense AGN jets, motivated by the
observation of double-double radio galaxies.  We simulated both a homogeneous jet and two jets with a different
spine--sheath structure. We find that a full outburst cycle is naturally divided into four different phases.   
\halfskip \noindent
Phase 1 lasts as long as the first jet is driven by the AGN. It can be characterized by:\\
- A jet-head advance speed that is very slow, \mbox{$v_{\rm hd} \ll c$};\\
- A strong bow shock and a strong Mach disc at the jet-head; \\
- Prominent high-pressure hotspots that remain visible throughout the entire phase; \\
- A strong back-flow of shocked jet material that collects in a thick cocoon; \\
- A rapid loss of radial integrity in the case of the isochoric jet.
\halfskip \noindent
Phase 2 occurs after the first jet is switched off, and no fresh jet material enters the system. In phase 2 we
find that:\\ 
- The remaining front-end of the initial jet continues to propagate towards the jet-head after the jet has been
switched off;\\
- In spine--sheath jets the jet spine outruns the jet sheath at the trailing end of the jet. Patches of material
originating from the jet sheath are left behind along the old jet path;\\
% The void that is left behind is filled up with sheath material;\\
- As soon as the trailing end of the first jet has crossed the Mach disc, the high-pressure hotspots disappear.
\halfskip \noindent
In phase 3 a new jet launches into the remnant cocoon of the initial jet. For this phase we find:\\
- A jet-head advance speed close to the jet bulk flow speed ($\sim c$);\\
- Significantly weaker bow shock and Mach disc near the jet-head. \\
- A much smaller mass deposition through the Mach disc for given jet length: a combined result of the larger jet
advance speed and relativistic time dilatation;\\
- The absence of prominent hotspots at the jet-head; \\
- A thin cocoon and, as a consequence, only a few weak vortices that are not capable of driving shocks into the
jet to promote mixing. Therefore, the jets retain their radial integrity, particularly for the case of the
isochoric jet.
\halfskip \noindent
Finally, phase 4 begins when the restarted jet breaks out of the older cocoon and then propagates further into
the undisturbed ambient medium. In this last phase we find:\\
- Propagation of the jet-head proceeds in a very similar fashion as in the first phase. \\
- Renewed formation of strong shocks and of a high-pressure hotspot at the head of the jet;\\
- A strong back-flow is re-established as soon as the restarted jet breaks out of the old cocoon. \\
- Most of the material from the first jet eruption is pushed outwards from the second jet-head and backwards
towards the parent galaxy. A very small fraction \mbox{$(<0.1\%)$} remains in the regions where the back-flow
driven by the restarted jet is strong.\\
%\nskip

The most prominent and distinctive phase is phase 3: its features closely resemble the observed properties and
morphology of double-double radio galaxies. 

\section*{Acknowledgments}
This research is funded by the {\it Nederlandse Onderzoekschool Voor Astronomie} (NOVA).
RK acknowledges funding from projects GOA/2009/009 (KU Leuven), G.0238.12 (FWO-Vlaanderen), BOF F+ financing
related to EC FP7/2007-2013 grant agreement SWIFF (no.263340), and the Interuniversity Attraction Poles Programme
initiated by the Belgian Science Policy Office (IAP P7/08 CHARM). OP is supported by STFC under the standard
grant ST/I001816/1. The authors acknowledge fruitful discussions with and coding efforts by Z. Meliani.

%S.M. also acknowledges support from The European Communities Seventh Framework Programme (FP7/2007-2013)
%under grant agreement number ITN 215212 “Black Hole Universe”.
%\end{comment}

%\clearpage

\footnotesize{
\bibliographystyle{./mn2e} % style file xxx.bst.
%\bibliography{/Users/sanderwalg/Work/Research-PhD/Tex-PhD/References.bib}
\bibliography{./2013References.bib}
}

%\clearpage

\bsp

%\appendix

%\section[]{App A}\label{sec:AppA}

\begin{comment}
\begin{figure}
%\begin{center}
%$
%\begin{array}{cc}
%\includegraphics[clip=false,width=0.5\textwidth,angle=0]{pics/figx.eps} \\
%\includegraphics[clip=true,trim=8cm 4.5cm 6.5cm 4cm,width=0.5\textwidth,angle=0]{pics/SimA2-B1545-321.png} \\
\includegraphics[clip=true,trim=0cm 0cm 0cm 0cm,width=0.5\textwidth,angle=0]
{pics/plotsX2/hllc/SimB1545-321-A2-145-t78-em4.eps} \\
%\end{array}
%$
%\end{center}

\caption{Synthesis map of the emitted synchrotron emission for the isochoric jet $A2$ using a Gaussian smoothing
with \mbox{FWHM = $5 \times 5$ kpc$^2$}. The horizontal and vertical scales are given in kpc. For comparison,
the original DDRG B1545-321 is shown at the bottom right, where \mbox{$1\arcsec \approx 2$ kpc}. The color map
shows the square root of the intensity. The image is constructed by embedding the 2.5$D$ simulations into a 3$D$
grid and integrating the synchrotron emissivity along the line of sight. The viewing angle is
\mbox{$\theta = 78 ^\circ$}, with the North-East part of the source pointing away from the observer. A spectral
index of $\alpha=0.7$ was used, similar to that inferred from B1545-321. The observation frequency in this
synchrotron emission model enters the emissivity as a constant. Since we work in arbitrary units, we set
\mbox{$\nu = 1$}. However, we impose the synchrotron cooling time at the chosen frequency large compared to the
adiabatic expansion time-scale.
}
  \label{fig:SimB1545-321}
\end{figure}
\end{comment}

\label{lastpage}

\end{document}